\documentclass[]{llncs}
\usepackage{cite}
\usepackage{url}
\usepackage[pdfusetitle,pdfauthor={Satoshi Kura, Hiroshi Unno, and Ichiro Hasuo}]{hyperref}
\usepackage{graphicx}
\usepackage{amsmath}
\usepackage{amssymb}
\usepackage{algorithm}
\usepackage{algorithmicx,algpseudocode}

\usepackage{mathtools}
\mathtoolsset{showonlyrefs}
\usepackage{color}
\usepackage{tikz,pgfplots}
\usetikzlibrary{shapes}
\usepackage{comment}
\usepackage{listings}
\usepackage[subrefformat=parens]{subcaption}
\usepackage{graphicx}
\usepackage{enumerate}
\usepackage{wrapfig}
\usepackage[firstpage]{draftwatermark}

\SetWatermarkText{\hspace*{5in}\raisebox{5.in}{\includegraphics[scale=0.8]{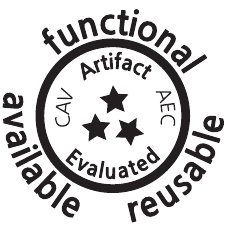}}}
\SetWatermarkAngle{0}

\usepackage{ifthen}
\newboolean{longversion}
\setboolean{longversion}{true} 
\newcommand{\referappendix}[2]{\ifthenelse{\boolean{longversion}}{Appendix~\ref{#1}}{\cite[Appendix {#2}]{arxiv}}}

\usepackage{macros}

\title{Decision Tree Learning \\in CEGIS-Based Termination Analysis}
\author{Satoshi Kura\inst{1,2}
\and Hiroshi Unno\inst{3,4}
\and Ichiro Hasuo\inst{1,2}}
\institute{National Institute of Informatics, Tokyo, Japan
\and The Graduate University for Advanced Studies (SOKENDAI), \\Kanagawa, Japan
\and University of Tsukuba, Ibaraki, Japan
\and RIKEN AIP, Tokyo, Japan}

\begin{document}

\maketitle

\begin{abstract}
We present a novel decision tree-based synthesis algorithm of ranking functions for verifying program termination.
Our algorithm is integrated into the workflow of CounterExample Guided Inductive Synthesis (CEGIS). CEGIS is an iterative learning model where, at each iteration, (1) a synthesizer synthesizes a candidate solution from the current examples, and (2) a validator accepts the candidate solution if it is correct, or rejects it providing counterexamples as part of the next examples. Our main novelty is in the design of a synthesizer: building on top of a usual decision tree learning algorithm, our algorithm detects \emph{cycles} in a set of example transitions and uses them for refining decision trees.
We have implemented the proposed method and obtained promising experimental results on existing benchmark sets of (non-)termination verification problems that require synthesis of piecewise-defined lexicographic affine ranking functions.
\end{abstract}

\section{Introduction}
\label{sec:intro}

\paragraph{Termination Verification by Ranking Functions and CEGIS}
Termination verification is a fundamental but challenging problem in program analysis.  Termination verification usually involves some well-foundedness arguments. Among them are those methods which synthesize \emph{ranking functions}~\cite{Floyd1967}: a ranking function assigns a natural number (or an ordinal, more generally) to each program state, in such a way that the assigned values strictly decrease along transition. Existence of such a ranking function witnesses termination, where well-foundedness of the set of natural numbers (or ordinals) is crucially used.

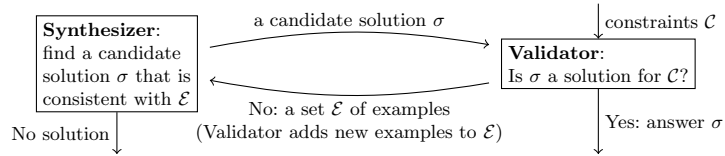
\begin{figure}[tbp]
	\centering
	\begin{tikzpicture}[scale=0.8, transform shape]
		\node[draw, align=left
	    ] at (0, 0) (synth) {\textbf{Synthesizer}:\\find a candidate \\ solution $\sigma$ that is\\consistent with $\examples$};
		\node[draw, align=left
	 ] at (8, 0) (valid) {\textbf{Validator}:\\ Is $\sigma$ a solution for $\clauses$?};
		\draw[->, shorten <=0.5em, shorten >=0.5em] (synth) edge[bend left=10] node[above,align=center] {a candidate  solution $\sigma$} (valid);
		\draw[->, shorten <=0.5em, shorten >=0.5em] (valid) edge[bend left=10] node[below,align=center] {No: a set $\examples$ of examples\\ (Validator adds new examples to $\examples$)} (synth);
		\draw[->] (valid) -- node[right] {Yes: answer $\sigma$} (8, -1.5);
		\draw[->] (8, 1) -- node[right] {constraints $\clauses$} (valid);
		\draw[->] (synth) -- node[left] {No solution} (0, -1.5);
	\end{tikzpicture}
	\caption{the CEGIS architecture}
	\label{fig:cegis}
\end{figure}

We study synthesis of ranking functions by 
CounterExample Guided Inductive Synthesis (CEGIS)~\cite{Solar-Lezama2006}. CEGIS is an iterative learning model in which a synthesizer and a validator interact to find solutions for given constraints.
At each iteration, (1) a synthesizer tries to find a candidate solution from the current examples, and (2) a validator accepts the candidate solution if it is correct, or rejects it providing counterexamples. These counterexamples are then used as part of the next examples (Fig.~\ref{fig:cegis}).

%
CEGIS has been applied not only to program verification tasks (synthesis of inductive invariants~\cite{Padhi2016,Padhi2019,Garg2014,Garg2016}, that of ranking functions~\cite{Gonnord2015}, etc.) but also to constraint solving (for CHC~\cite{Champion2018,Ezudheen2018,Zhu2018,Satake2020}, for \PCSPWF{}($\theory$)~\cite{Unno2020,Unno2021}, etc.). The success of CEGIS is attributed to the degree of freedom that synthesizers enjoy. In CEGIS, synthesizers receive a set of individual examples that synthesizers can use in various creative and speculative manners (such as machine learning). In contrast,  in other methods such as~\cite{Podelski2004,Alias2010,BenAmram2014,Leike2014a,Ben-Amram2017,BenAmram2019}, synthesizers receive logical constraints that are much more binding.

\paragraph{Segmented Synthesis in CEGIS-Based Termination Analysis}
The choice of a \emph{candidate space} for candidate solutions $\sigma$
 is important in CEGIS.
A candidate space should  be \emph{expressive}: by limiting a candidate space, the CEGIS architecture may miss a genuine solution.
At the same time, \emph{complexity} should be low: a larger candidate space tends to be more expensive for synthesizers to handle.

 This tradeoff is also in the choice of the type of examples: using an expressive example type, a small number of examples can prune a large portion of the candidate space; however, finding such expressive examples
 tends to be expensive.


In this paper, we use \emph{piecewise affine functions} as our candidate space for ranking functions.
Piecewise affine functions are functions of the form
\begin{equation}\label{eq:introPiecewiseAffine}
  f(\seq{x}) \quad=\quad \begin{cases}
	\seq{a}_1 \cdot \seq{x} + b_1 & \seq{x} \in L_1 \\
	\quad\vdots \\
	\seq{a}_n \cdot \seq{x} + b_n & \seq{x} \in L_n
\end{cases}  
\end{equation}
where $\{ L_1, \dots, L_n \}$ is a partition of the domain of $f(\seq{x})$ such that each $L_i$ is a polyhedron (i.e.\ a conjunction of linear inequalities).
We say \emph{segmented synthesis} to emphasize that our synthesis targets are piecewise affine functions with case distinction. Piecewise affine functions stand on a good balance between expressiveness and complexity: the tasks of synthesizers and validators can be reduced to linear programming (LP);
at the same time, case distinction allows them to model a variety of situations, especially where there are discontinuities in the function values and/or derivatives.

We use \emph{transition examples} as our example type (Table~\ref{tab:ranking_function_cegis}).
Transition examples are pairs of program states that represent transitions; they are much cheaper to handle compared to \emph{trace examples} (finite traces of executions until termination) used e.g.\ in~\cite{Urban2016,Fedyukovich2018}.
The current work is the first to pursue segmented synthesis of ranking functions with transition examples; see Table~\ref{tab:ranking_function_cegis}.

\begin{table}[tbp]
	\centering
	\caption{ranking function synthesis by CEGIS}
	\label{tab:ranking_function_cegis}
	\begin{tabular}{c|c@{\hspace{1.5em}}c}
		candidate space $\setminus$ example type & trace examples & transition examples \\
		\hline
		affine ranking functions & \cite{Urban2016,Fedyukovich2018} & \cite{Gonnord2015} \\
		piecewise affine ranking functions & \cite{Urban2016,Fedyukovich2018} & our method
	\end{tabular}
\end{table}

\paragraph{Decision Tree Learning for CEGIS-Based Termination Analysis: a Challenge}
In this paper, we represent piecewise affine functions~\eqref{eq:introPiecewiseAffine} by the data structure of \emph{decision trees}. The data structure suits the CEGIS architecture (Fig.~\ref{fig:cegis}): iterative refinement of candidate solutions can be naturally expressed by growing decision trees.
The main challenge of this paper is the design of an effective synthesizer for decision trees---such a synthesizer \emph{learns} decision trees from examples.

In fact,  decision tree learning in the CEGIS architecture has already been actively pursued, for the synthesis of \emph{invariants} as opposed to ranking functions~\cite{Krishna2015,Garg2016,Champion2018,Ezudheen2018,Zhu2018}. It is therefore a natural idea to adapt the decision tree learning algorithms used there, from invariants to ranking functions. However, we find that a naive adaptation of those algorithms for invariants does not suffice: they are good at handling \emph{state examples} that appear in CEGIS for invariants; but they are not good at handling transition examples.

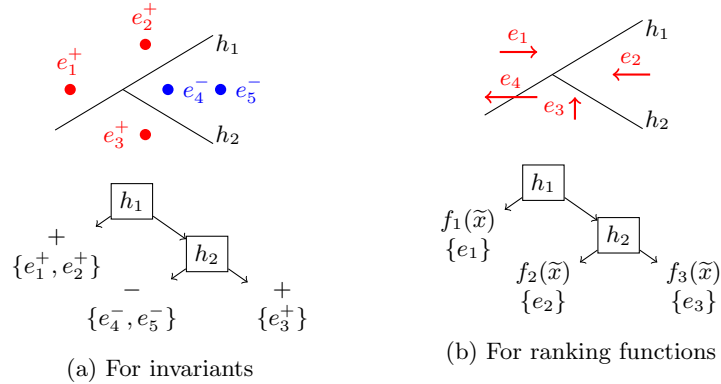
\begin{figure}[tbp]
	\centering
	\begin{minipage}{0.45\linewidth}
		\centering
		\begin{tikzpicture}[yscale=0.6]
			\node[red,label={[red]above:$e_1^{+}$},circle,fill,inner sep=1.5pt] at (-1, 0) {};
			\node[red,label={[red]above:$e_2^{+}$},circle,fill,inner sep=1.5pt] at (0, 1) {};
			\node[red,label={[red]left:$e_3^{+}$},circle,fill,inner sep=1.5pt] at (0, -1) {};
			\node[blue,label={[blue]right:$e_4^{-}$},circle,fill,inner sep=1.5pt] at (0.3, 0) {};
			\node[blue,label={[blue]right:$e_5^{-}$},circle,fill,inner sep=1.5pt] at (1, 0) {};
			\draw (-1.2, -0.9) -- (0.9, 1.2);
			\draw (-0.3, 0) -- (0.9, -1.2);
			\node at (1.1, 1.0) {$h_1$};
			\node at (1.1, -1.0) {$h_2$};
		\end{tikzpicture} \\[1em]
		\begin{tikzpicture}[yscale=0.7]
			\node[draw] (h1) at (0, 0) {$h_1$};
			\node[draw] (h2) at (1, -1) {$h_2$};
			\draw[->] (h1) -- (h2);
			\node[align=center] (l1) at (-1, -1) {+ \\ $\{ e_1^{+}, e_2^{+} \}$};
			\draw[->,shorten >=-1ex] (h1) -- (l1);
			\node[align=center] (l2) at (0, -2) {$-$ \\ $\{ e_4^{-}, e_5^{-} \}$};
			\node[align=center] (l3) at (2, -2) {+ \\ $\{ e_3^{+} \}$};
			\draw[->,shorten >=-1ex] (h2) -- (l2);
			\draw[->] (h2) -- (l3);
		\end{tikzpicture}
		\subcaption{For invariants}
		\label{fig:decision_tree_learning:invariant}
	\end{minipage}
	\begin{minipage}{0.45\linewidth}
		\centering
		\begin{tikzpicture}[yscale=0.6]
			\draw (-1.2, -0.9) -- (0.9, 1.2);
			\draw (-0.3, 0) -- (0.9, -1.2);
			\node at (1.1, 1.0) {$h_1$};
			\node at (1.1, -1.0) {$h_2$};
			\draw[->, red, thick] (-1, 0.5) -- node[above] {$e_1$} (-0.5, 0.5);
			\draw[->, red, thick] (1, 0) -- node[above] {$e_2$} (0.5, 0);
			\draw[->, red, thick] (0, -1) -- node[left] {$e_3$} (0, -0.5);
			\draw[->, red, thick] (-0.5, -0.5) -- node[above] {$e_4$} (-1.2, -0.5);
		\end{tikzpicture} \\[1em]
		\begin{tikzpicture}[yscale=0.7]
			\node[draw] (h1) at (0, 0) {$h_1$};
			\node[draw] (h2) at (1, -1) {$h_2$};
			\draw[->] (h1) -- (h2);
			\node[align=center] (l1) at (-1, -1) {$f_1(\seq{x})$ \\ $\{ e_1 \}$};
			\draw[->] (h1) -- (l1);
			\node[align=center] (l2) at (0, -2) {$f_2(\seq{x})$ \\ $\{ e_2 \}$};
			\node[align=center] (l3) at (2, -2) {$f_3(\seq{x})$ \\ $\{ e_3 \}$};
			\draw[->] (h2) -- (l2);
			\draw[->] (h2) -- (l3);
		\end{tikzpicture}
		\subcaption{For ranking functions}
		\label{fig:decision_tree_learning:ranking}
	\end{minipage}
	\caption{Decision tree learning}
	\label{fig:decision_tree_learning}
\end{figure}


%
%
More specifically, when  decision tree learning is applied to invariant synthesis (Fig.~\ref{fig:decision_tree_learning:invariant}), examples are given in the form of program states labeled as positive or negative.
Decision trees are then built by iteratively selecting the best halfspaces---where ``best'' is in terms of some quality measures---until each leaf contains examples with the same label. One common quality measure used here is an information-theoretic notion of \emph{information gain}.

We extend this from invariant synthesis to ranking function synthesis where examples are given by transitions instead of states (Fig.~\ref{fig:decision_tree_learning:ranking}). In this case, a major challenge is to cope with examples that cross a border of the current segmentation---such as the transition $e_4$ crossing the border $h_1$ in Fig.~\ref{fig:decision_tree_learning:ranking}. 
Our decision tree learning algorithm should handle such crossing examples, taking into account the constraints imposed on the leaf labels  affected by those examples (the affected leaf labels are $f_1(\seq{x})$ and $f_3(\seq{x})$ in the case of $e_4$).

\paragraph{Our Algorithm: Cycle-Based Decision Tree Learning for Transition Examples}
%
We use what we call the \emph{cycle detection theorem} (Theorem~\ref{thm:sufficient_condition_decision_tree_lexicographic}) as a theoretical tool to handle such crossing examples.
The theorem claims the following: if there is no piecewise affine ranking function with the current segmentation of the domain (such as the one in Fig.~\ref{fig:decision_tree_learning:ranking} given by  $h_1$ and $h_2$), then this must be caused by a certain type of cycle of constraints, which we call an \emph{implicit cycle}.

In our decision tree learning algorithm, when we do not find a piecewise affine ranking function with the current segmentation, we find an implicit cycle and refine the segmentation to break the cycle. Once all the implicit cycles are gone, the cycle detection theorem guarantees the existence of a candidate piecewise affine ranking function with the segmentation.



We integrate this decision tree learning algorithm in the CEGIS architecture (Fig.~\ref{fig:cegis}) and use it as a synthesizer. Our implementation of this framework
gives promising experimental results on existing benchmark sets.

\paragraph{Contribution}
Our contribution is summarized as follows.
\begin{itemize}
	\item We provide a decision tree-based synthesizer for ranking functions integrated into the CEGIS architecture. Our synthesizer uses transition examples to find candidate piecewise affine ranking functions. A major challenge here, namely handling constraints arising from crossing examples, is coped with by our theoretical observation of the cycle detection theorem.
	\item We implement our synthesizer for ranking functions implemented in \muval{} and report the experience of using \muval{} for termination and non-termination analysis. The experiment results show that \muval{}'s performance is comparable to state-of-the-art termination analyzers~\cite{Emmes2012, Brockschmidt2012,BenAmram2014,Heizmann2014} from Termination Competition 2020, and that \muval{} can prove \mbox{(non-)}ter\-mination of some benchmarks with which other analyzers struggle.
\end{itemize}

\paragraph{Organization.}
Section~\ref{sec:overview} shows the overview of our method via examples.
Section~\ref{sec:problem} explains our target class of predicate constraint satisfaction problems and how to encode (non-)termination problem into such constraints.
In Section~\ref{sec:cegis}, we review CEGIS architecture, and then explain simplification of examples into positive/negative examples.
Section~\ref{sec:synth_rank} proposes our main contribution, our decision tree-based ranking function synthesizer.
Section~\ref{sec:eval} shows our implementation and experimental results.
Related work is discussed in Section~\ref{sec:related}, and we conclude in Section~\ref{sec:conc}.

\section{Preview by Examples}
\label{sec:overview}
We present a preview of our method using concrete examples. We start with an overview of the general CEGIS architecture, after which we proceed to our main contribution, namely a decision tree learning algorithm for transition examples.

\subsection{Termination Verification by CEGIS}\label{subsec:overview_cegis}
Our method follows the usual workflow of termination verification by CEGIS. It works as follows:
given a program, we encode the termination problem into a constraint solving problem, and then use the CEGIS architecture to solve the constraint solving problem.

\paragraph{Encoding the termination problem.}
The first step of our method is to encode the termination problem as the set $\clauses$ of constraints.

\begin{myexample}
As a running example, consider the following C program.
\begin{lstlisting}[language={C},basicstyle={\small\ttfamily},tabsize = 4]
while(x != 0) { if(x < 0) { x++; } else { x--; } }
\end{lstlisting}
The termination problem is encoded as the following constraints.
\begin{align}
	x < 0 \land x' = x + 1 &\implies R(x, x') \label{eq:wf_true_branch} \\
	\lnot (x < 0) \land x' = x - 1 &\implies R(x, x') . \label{eq:wf_false_branch}
\end{align}
Here, $R$ is a predicate variable representing a well-founded relation, and term variables $x, x'$ are universally quantified implicitly.
\end{myexample}

The set $\clauses$ of constraints claims that the transition relation for the given program is subsumed by a well-founded relation.
So, verifying termination is now rephrased as the existence of a solution for $\clauses$.
Note that we omitted constraints for invariants for simplicity in this example (see Sect.~\ref{sec:problem} for the full encoding).

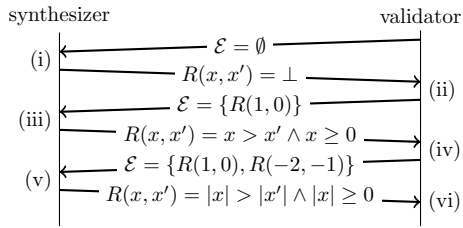
\begin{wrapfigure}{r}{19em}
	\centering
	\vspace*{-2em}
\begin{tikzpicture}[scale=0.8, transform shape]
	\node (synth) at (-3, 0) {synthesizer};
	\draw (synth) -- (-3, -3.5);
	\node (valid) at (3, 0) {validator};
	\draw (valid) -- (3, -3.5);
	\draw[->, thick] (3, -0.4) -- node[fill=white] {$\examples = \emptyset$} (-3, -0.6);
	\node[left] at (-3, -0.75) {(i)};
	\draw[->, thick] (-3, -0.9) -- node[fill=white] {$R(x, x') = \bot$} (3, -1.1);
	\node[right] at (3, -1.25) {(ii)};
	\draw[->, thick] (3, -1.4) -- node[fill=white] {$\examples = \{R(1, 0)\}$} (-3, -1.6);
	\node[left] at (-3, -1.75) {(iii)};
	\draw[->, thick] (-3, -1.9) -- node[fill=white] {$R(x, x') = x > x' \land x \ge 0$} (3, -2.1);
	\node[right] at (3, -2.25) {(iv)};
	\draw[->, thick] (3, -2.4) -- node[fill=white] {$\examples = \{ R(1, 0), R(-2, -1) \}$} (-3, -2.6);
	\node[left] at (-3, -2.75) {(v)};
	\draw[->, thick] (-3, -2.9) -- node[fill=white] {$R(x, x') = |x| > |x'| \land |x| \ge 0$} (3, -3.1);
	\node[right] at (3, -3.1) {(vi)};
\end{tikzpicture}
\caption{An example of CEGIS iterations}
\label{fig:cegis_iterations}
\vspace*{-2em}
\end{wrapfigure}
\paragraph{Constraint solving by CEGIS.}
The next step is to solve $\clauses$ by CEGIS.

In the CEGIS architecture, a synthesizer and a validator iteratively exchange a set $\examples$ of examples and a candidate solution $R(x, x')$ for $\clauses$.
At the moment, we present a rough sketch of CEGIS, leaving the details of our implementation to Sect.~\ref{subsec:handling_cycles}.

\begin{myexample}
Fig.~\ref{fig:cegis_iterations} shows how the CEGIS architecture solves the set $\clauses$ of constraints shown in \eqref{eq:wf_true_branch} and \eqref{eq:wf_false_branch}.
Fig.~\ref{fig:cegis_iterations} consists of three pairs of interactions (i)-(vi) between a synthesizer and a validator.
\begin{enumerate}[(i)]
	\item The synthesizer takes $\examples = \emptyset$ as a set of examples and returns a candidate solution $R(x, x') = \bot$ synthesized from $\examples$. In general, candidate solutions are required to satisfy all constraints in $\examples$, but the requirement is vacuously true in this case.
	\item The validator receives the candidate solution and finds out that the candidate solution is not a genuine solution. The validator finds that the assignment $x = 1, x' = 0$ is a counterexample for \eqref{eq:wf_false_branch}, and thus adds $R(1, 0)$ to $\examples$ to prevent the same candidate solution in the next iteration.
	\item The synthesizer receives the updated set $\examples = \{ R(1, 0) \}$ of examples, finds a ranking function $f(x) = x$ for $\examples$ (i.e.\ for the transition from $x = 1$ to $x' = 0$), and returns a candidate solution $R(x, x') = x > x' \land x \ge 0$.
	\item The validator checks the candidate solution, finds a counterexample $x = -2, x' = -1$ for \eqref{eq:wf_true_branch}, and adds $R(-2, -1)$ to $\examples$.
	\item The synthesizer finds a ranking function $f(x) = |x|$ for $\examples$ and returns $R(x, x') = |x| > |x'| \land |x| \ge 0$ as a candidate solution. Note that the synthesizer have to synthesize a piecewise affine function here, but details are deferred to Sect.~\ref{subsec:handling_cycles}.
	\item The validator accepts the candidate solution because it is a genuine solution for $\clauses$.
\end{enumerate}
\end{myexample}

\subsection{Handling Cycles in Decision Tree Learning}\label{subsec:handling_cycles}
\begin{wrapfigure}{r}{15em}
	\centering
	\vspace*{-2em}
	\begin{tikzpicture}[scale=0.8, transform shape]
		\node[draw] (n1) at (0, 0) {$y \ge 0$?};
		\node[draw] (n2) at (-1.4, -1.5) {$x-1 \ge 0$?};
		\node[inner sep=1pt, outer sep=1pt] (l1) at (1.4, -1.5) {$f(x, y) = -y$};
		\node[inner sep=1pt, outer sep=1pt] (l2) at (-2.8, -3) {$f(x, y) = x-1$};
		\node[inner sep=1pt, outer sep=1pt] (l3) at (0, -3) {$f(x, y) = 1-x$};
		\draw[thick, ->] (0, 0.7) -- (n1);
		\draw[thick, ->] (n1) -- (l1) node[midway, above right] {$<$};
		\draw[thick, ->] (n1) -- (n2) node[midway, above left] {$\ge$};
		\draw[thick, ->] (n2) -- (l2) node[midway, above left] {$\ge$};
		\draw[thick, ->] (n2) -- (l3) node[midway, above right] {$<$};
	\end{tikzpicture}
	\begin{minipage}[b]{15em}
		\begin{align}
			&f(x, y) = \\
			&\begin{cases}
				x - 1 & y \ge 0 \land x - 1 \ge 0 \\
				1 - x & y \ge 0 \land x - 1 < 0 \\
				-y & y < 0
			\end{cases}
		\end{align}
	\end{minipage}
	\caption{An example of a decision tree that represents a piecewise affine ranking function $f(x, y)$}
	\label{fig:example_decision_tree}
	\vspace*{-3.5em}
\end{wrapfigure}

We explain the importance of handling cycles in our decision tree-based synthesizer of piecewise affine ranking functions.

In what follows, we deal with such decision trees as shown in Fig.~\ref{fig:example_decision_tree}: their internal nodes have affine inequalities (i.e.\ halfspaces); their leaves have affine functions; and overall, such a decision tree expresses a piecewise affine function (Fig.~\ref{fig:example_decision_tree}). When we remove leaf labels from such a decision tree, then we obtain a template of piecewise functions where condition guards are given but function bodies are not. We shall call the latter a \emph{segmentation}.

\paragraph{Input and output of our synthesizer.}
The input of our synthesizer is a set $\examples$ of transition examples (e.g.\ $\examples = \{R(1, 0), R(-2, -1)\}$) as explained in Sect.~\ref{subsec:overview_cegis}.
The output of our synthesizer is a well-founded relation $R(\seq{x}, \seq{x}') \coloneqq f(\seq{x}) > f(\seq{x}') \land f(\seq{x}) \ge 0$ where $\seq{x}$ is a sequence of variables and $f(\seq{x})$ is a piecewise affine function, which is represented by a decision tree (Fig.~\ref{fig:example_decision_tree}). Therefore our synthesizer aims at \emph{learning} a suitable decision tree.

\paragraph{Refining segmentations and handling cycles.}
Roughly speaking, our synthesizer learns decision trees in the following steps. 
\begin{enumerate}
	\item Generate a set $H$ of halfspaces from the given set $\examples$ of examples. 
	This $H$ serves as the vocabulary for internal nodes.
Set the initial segmentation to be the one-node tree (i.e.\ the trivial segmentation). 
	\item Try to synthesize a piecewise affine ranking function $f$ for $\examples$ with the current segmentation---that is,  try to find suitable leaf labels. If found, then use this $f$ in a candidate well-founded relation $R(\seq{x}, \seq{x}') = f(\seq{x}) > f(\seq{x}') \land f(\seq{x}) \ge 0$.\label{enum:learning_step2}
	\item Otherwise, refine the current segmentation with some halfspace in $H$, and go  to Step~\ref{enum:learning_step2}.\label{enum:learning_step3}
\end{enumerate}
The key step of our synthesizer is Step~\ref{enum:learning_step3}.
We show a few examples.

\begin{figure}[tbp]
	\centering
	\begin{minipage}[b]{0.48\linewidth}
		\centering
		\begin{tikzpicture}
			\begin{axis}[
				axis x line=middle,
				axis y line=none,
				domain=-2.5:2.5,
				xtick={-2, -1, 0, 1, 2},
				samples=1001,
				height=0.5\textwidth,
				width=\textwidth,
				yscale=0.6,
			]
				\draw[->, very thick, red] (axis cs:-1.95, -0.2) -- (axis cs:-1.05, -0.2);
				\draw[->, very thick, red] (axis cs:0.95, -0.2) -- (axis cs:0.05, -0.2);
				\addplot [mark=none, dashed] coordinates {(0, -0.5) (0, 2.5)};
				\addplot[no marks] {abs(x)};
				\node at (axis cs:2, 1.3) {$f(x)$};
			\end{axis}
		\end{tikzpicture}
		\subcaption{Good ($x \ge 0$)}
		\label{fig:select_halfspaces:xge0}
	\end{minipage}
	\begin{minipage}[b]{0.48\linewidth}
		\centering
		\begin{tikzpicture}
			\begin{axis}[
				axis x line=middle,
				axis y line=none,
				domain=-2.5:2.5,
				xmin=-2.5,
				xmax=2.5,
				xtick={-2, -1, 0, 1, 2},
				samples=1001,
				height=0.5\textwidth,
				width=\textwidth,
				yscale=0.6,
			]
				\draw[->, very thick, red] (axis cs:-1.95, -0.2) -- (axis cs:-1.05, -0.2);
				\draw[->, very thick, red] (axis cs:0.95, -0.2) -- (axis cs:0.05, -0.2);
				\addplot [mark=none, dashed] coordinates {(-2, -0.5) (-2, 2.5)};
			\end{axis}
		\end{tikzpicture}
		\subcaption{Bad ($x \ge -2$)}
		\label{fig:select_halfspaces:xp2ge0}
	\end{minipage}
	\caption{Selecting halfspaces. Transition  examples are shown by red arrows. Boundaries of halfspaces are shown by dashed lines.}
	\label{fig:select_halfspaces}
\end{figure}
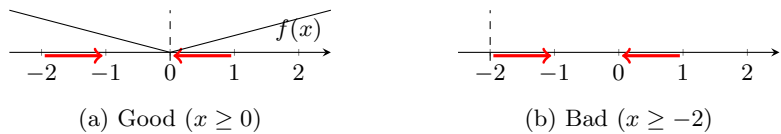

\begin{myexample}\label{ex:decision_tree_learning1}
	Suppose we are given $\examples = \{ R(1, 0), R(-2, -1) \}$ as a set of examples.
	Our synthesizer proceeds as follows:
	(1) Our synthesizer generates the set $H \coloneqq \{ x \ge 1, x \ge 0, x \ge -2, x \ge -1\}$ from the examples in $\examples$.
	(2) Our synthesizer tries to find a ranking function of the form $f(x) = a x + b$ (with the trivial segmentation), but there is no such ranking function.
	(3) Our synthesizer refines the current segmentation with $(x \ge 0) \in H$ because $x \ge 0$ ``looks good''.
	(4) Our synthesizer tries to find a ranking function of the form $f(x) = \mathbf{if}\ x \ge 0\ \mathbf{then}\ a x + b\ \mathbf{else}\ c x + d$, using the current segmentation.
	Our synthesizer obtains $f(x) = \mathbf{if}\ x \ge 0\ \mathbf{then}\ x\ \mathbf{else}\ -x$ and use this $f(x)$ for a candidate solution.

	How can we decide which halfspace in $H$ ``looks good''?
	We use \emph{quality measure} that is a value representing the quality of each halfspace and select the halfspace with the maximum quality measure.

	Fig.~\ref{fig:select_halfspaces} shows the comparison of the quality of $x \ge 0$ and $x \ge -2$ in this example.
	Intuitively, $x \ge 0$ is better than $x \ge -2$ because we can obtain a simple ranking function $\mathbf{if}\ x \ge 0\ \mathbf{then}\ x\ \mathbf{else}\ -x$ with $x \ge 0$ (Fig.~\ref{fig:select_halfspaces:xge0}) while we need further refinement of the segmentation with $x \ge -2$ (Fig.~\ref{fig:select_halfspaces:xp2ge0}).
	In Sect.~\ref{sec:synth_rank}, we introduce a quality measure for halfspaces following this intuition.

	Our synthesizer iteratively refines segmentations following this quality measure, until examples contained in each leaf of the decision tree admit an affine ranking function.
	This approach is inspired by the use of information gain in the decision tree learning for invariant synthesis.
\end{myexample}

Example~\ref{ex:decision_tree_learning1} showed a natural extension of a decision tree learning method for invariant synthesis.
However, this is not enough for transition examples, for the reasons of \emph{explicit} and \emph{implicit cycles}. Here are their examples.

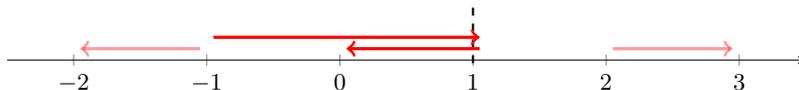
\begin{figure}[tbp]
	\centering
	\begin{tikzpicture}
		\begin{axis}[
			axis x line=middle,
			axis y line=none,
			xmin=-2.5,
			xmax=3.5,
			ymin=-0.2,
			ymax=1.5,
			xtick={-2, -1, 0, 1, 2, 3},
			height=0.2\textwidth,
			width=\textwidth,
		]
			\addplot [mark=none, thick, dashed] coordinates {(1, -0.5) (1, 1.5)};	
			\draw[->, very thick, red] (axis cs:-0.95, 0.6) -- (axis cs:1.05, 0.6);
			\draw[->, very thick, red] (axis cs:1.05, 0.3) -- (axis cs:0.05, 0.3);
			\draw[->, very thick, red, opacity=0.4] (axis cs:-1.05, 0.3) -- (axis cs:-1.95, 0.3);
			\draw[->, very thick, red, opacity=0.4] (axis cs:2.05, 0.3) -- (axis cs:2.95, 0.3);
		\end{axis}
	\end{tikzpicture}
	\caption{Two examples $R(-1, 1)$ and $R(1, 0)$ make an implicit cycle between $x \ge 1$ and $\lnot (x \ge 1)$.}
	\label{fig:phase3}
\end{figure}

\begin{myexample}\label{ex:decision_tree_learning2} 
	Suppose we are given $\examples = \{ R(1, 0), R(0, 1) \}$.
	In this case, there is no ranking function because $\examples$ contains a cycle $1 \to 0 \to 1$ witnessing non-termination.
	We call such a cycle an \emph{explicit cycle}.
\end{myexample}
\begin{myexample}\label{ex:decision_tree_learning3}
 Let $\examples = \{ R(-1, 1), R(1, 0), R(-1, -2), R(2, 3) \}$ (Fig.~\ref{fig:phase3}). 
	Our synthesizer proceeds as follows.
	(1) Our synthesizer generates the set $H \coloneqq \{ x \ge 1, x \ge 0, \dots\}$ of halfspaces.
	(2) Our synthesizer tries to find a ranking function of the form $f(x) = a x + b$ (with the trivial segmentation), but there is no such.
	(3) Our synthesizer refines the current segmentation with $(x \ge 1) \in H$ because $x \ge 1$ ``looks good'' (i.e.\ is the best with respect to a quality measure).

	We have reached the point where the naive extension of decision tree learning explained in Example~\ref{ex:decision_tree_learning1} no longer works: although all constraints contained in each leaf of the decision tree admit an affine ranking function, there is no piecewise affine ranking function for $\examples$ of the form $f(x) = \mathbf{if}\ x \ge 1\ \mathbf{then}\ a x + b\ \mathbf{else}\ c x + d$.

	More specifically, in this example, the leaf representing $x \ge 1$ contains $R(2, 3)$, and the other leaf representing $\lnot (x \ge 1)$ contains $R(-1, -2)$.
	The example $R(2, 3)$ admits an affine ranking function $f_1(x) = - x + 2$, and $R(-1, -2)$ admits $f_2(x) = x + 1$, respectively.
	However, the combination $f(x) = \mathbf{if}\ x \ge 1\ \mathbf{then}\ f_1(x)\ \mathbf{else}\ f_2(x)$ is not a ranking function for $\examples$.
	Moreover, there is no ranking function for $\examples$ of the form $f(x) = \mathbf{if}\ x \ge 1\ \mathbf{then}\ a x + b\ \mathbf{else}\ c x + d$.

	It is  clear that this failure is caused by the \emph{crossing examples}
        $R(-1, 1)$ and $R(1, 0)$.
        It is not that every crossing example is harmful. However, in this case, the set $\{ R(-1, 1), R(1, 0) \}$ forms a cycle between the leaf for $x \ge 1$ and the leaf for $\lnot (x \ge 1)$ (see Fig.~\ref{fig:phase3}). This ``cycle'' among leaves---in contrast to \emph{explicit} cycles such as $\{ R(1, 0), R(0, 1) \}$ in Example~\ref{ex:decision_tree_learning2}---is called an \emph{implicit cycle}.

	Once an implicit cycle is found, our synthesizer cuts it by refining the current segmentation.
	Our synthesizer continues the above steps (1--3) of decision tree learning as follows.
	(4) Our synthesizer selects $(x \ge 0) \in H$ and cuts the implicit cycle $\{ R(-1, 1), R(1, 0) \}$ by refining segmentations.
	(5) Using the refined segmentation, our synthesizer obtains $f(x) = \mathbf{if}\ x \ge 1\ \mathbf{then}\ - x + 2\ \mathbf{else}\ \mathbf{if}\ x \ge 0\ \mathbf{then}\ 0\ \mathbf{else}\ x + 3$ as a ranking function for $\examples$.
\end{myexample}

As explained in Example~\ref{ex:decision_tree_learning2},\ref{ex:decision_tree_learning3}, handling (explicit and implicit) cycles is crucial in decision tree learning for transition examples.
Moreover, our \emph{cycle detection theorem} (Theorem~\ref{thm:sufficient_condition_decision_tree_lexicographic}) claims that if there is no explicit or implicit cycle, then one can find a ranking function for $\examples$ without further refinement of segmentations.

\section{(Non-)Termination Verification as Constraint Solving}
\label{sec:problem}
We explain how to encode (non-)termination verification to constraint solving.

Following ~\cite{Unno2021}, we formalize our target class \PCSPWF{} of predicate constraint satisfaction problems parametrized by a first-order theory $\theory$.
\begin{mydefinition}
	Given a formula $\phi$, let $\ftv{\phi}$ be the set of free term variables and $\fpv{\phi}$ be the set of free predicate variables in $\phi$.
\end{mydefinition}
\begin{mydefinition}
	A \PCSPWF{} is defined as a pair $(\clauses,\WFRs)$ where
	$\clauses$ is a finite set of clauses of the form
	\begin{equation}
		\phi \lor \left( \bigvee_{i=1}^{\ell} X_i(\seq{t}_i) \right) \lor \left( \bigvee_{i=\ell+1}^m \neg X_i(\seq{t}_i) \right) \label{eq:pcspwf_clause}
	\end{equation}
	and $\WFRs \subseteq \fpv{\clauses}$ is a set of predicate variables that are required to denote \emph{well-founded} relations.
	Here, $0 \leq \ell \leq m$.
	Meta-variables $t$ and $\phi$ range over $\theory$-terms and $\theory$-formulas, respectively, such that $\ftv{\phi}=\emptyset$.
	Meta-variables $x$ and $X$ range over term and predicate variables, respectively.
\end{mydefinition}
A \PCSPWF{} $(\clauses,\WFRs)$ is called \CHCS{} (constrained Horn clauses,~\cite{Bjorner2015a}) if $\WFRs = \emptyset$ and $\ell \leq 1$ for all clauses $c \in \clauses$.  The class of \CHCS{} has been widely studied in the verification community~\cite{Champion2018,Ezudheen2018,Zhu2018,Satake2020}.

\begin{mydefinition}
	A \emph{predicate substitution} $\sigma$ is a finite map from predicate variables $X$ to closed predicates of the form $\lambda x_1,\dots,x_{\arity{X}}.\phi$.  We write $\dom{\sigma}$ for the domain of $\sigma$ and $\sigma(\clauses)$ for the application of $\sigma$ to $\clauses$.
\end{mydefinition}
\begin{mydefinition}
	A predicate substitution $\sigma$ is a \emph{(genuine) solution} for $(\clauses,\WFRs)$ if (1) $\fpv{\clauses} \subseteq \dom{\sigma}$; (2) $\models \bigwedge \sigma(\clauses)$ holds; and (3) for all $X \in \WFRs$, $\sigma(X)$ represents a well-founded relation, that is,  $\sort{\sigma(X)}=(\seq{s},\seq{s}) \to \propos$ for some sequence $\seq{s}$ of sorts and there is no infinite sequence $\seq{v}_1,\seq{v}_2,\dots$ of sequences $\seq{v}_i$ of values of the sorts $\seq{s}$ such that $\models \rho(X)(\seq{v}_i,\seq{v}_{i+1})$ for all $i\geq 1$.
\end{mydefinition}

\paragraph{Encoding termination.}
Given a set of initial state $\iota(\seq{x})$ and a transition relation $\tau(\seq{x}, \seq{x}')$, the termination verification problem is expressed by the \PCSPWF{} $(\clauses, \WFRs)$ where $\WFRs = \{ R \}$, and $\clauses$ consists of the following clauses.
\begin{align}
	\iota(\seq{x}) \implies I(\seq{x}) \qquad
	\tau(\seq{x}, \seq{x}') \land I(\seq{x}) \implies I(\seq{x}') \qquad
	\tau(\seq{x}, \seq{x}') \land I(\seq{x}) \implies R(\seq{x}, \seq{x}')
\end{align}
We use $\phi \implies \psi$ as syntax sugar for $\lnot \phi \lor \psi$, so this is a \PCSPWF{}.
The well-founded relation $R$ asserts that $\tau$ is terminating.
We also consider an invariant $I$ for $\tau$ to avoid synthesizing ranking functions on unreachable program states.

\paragraph{Encoding non-termination.}
We can also encode a problem of non-termination verification to \PCSPWF{} via recurrent sets~\cite{Gupta2008}.
For simplicity, we explain the encoding for the case of only one program variable $x$.
We consider a recurrent set $R$ satisfying the following conditions.
\begin{align}
	\iota(x) &\implies R(x) \label{eq:recurrent_initial} \\
	R(x) &\implies \exists x'. \tau(x, x') \land R(x') \label{eq:recurrent_exist}
\end{align}
To remove $\exists$ from~\eqref{eq:recurrent_exist}, we use the following constraint that is equivalent to~\eqref{eq:recurrent_exist}.
\begin{align}
	R(x) &\implies E(x, 0) \label{eq:recurrent_skolem1} \\
	E(x, x') &\implies \big(\tau(x, x') \land R(x')\big) \\
		&\lor \big(S(x', x'-1) \land E(x, x'-1)\big) 
		\lor \big(S(x', x'+1) \land E(x, x'+1)\big) \label{eq:recurrent_skolem2}
\end{align}

The intuition is as follows.
Given $x$ in the recurrent set $R$, the relation $E(x, x')$ searches for the value of $\exists x'$ in~\eqref{eq:recurrent_exist}.
The search starts from $x' = 0$ in~\eqref{eq:recurrent_skolem1}, and $x'$ is nondeterministically incremented or decremented in~\eqref{eq:recurrent_skolem2}.
The well-founded relation $S$ asserts that the search finishes within finite steps.
As a result, we obtain a \PCSPWF{} for non-termination defined by $(\clauses, \WFRs)$ where $\WFRs = \{ S \}$ and $\clauses$ is given by \eqref{eq:recurrent_initial}, \eqref{eq:recurrent_skolem1}, and (the disjunctive normal form of)~\eqref{eq:recurrent_skolem2}.
\begin{myexample}
	Consider the following C program.
	\begin{lstlisting}[language={C},basicstyle={\small\ttfamily},tabsize = 4]
	while(x > 0) { x = -2 * x + 9; }
	\end{lstlisting}
	The non-termination problem is encoded as the \PCSPWF{} $(\clauses, \WFRs)$ where $\WFRs = \{ S \}$, and $\clauses$ consists of
	\begin{align}
		x > 0 &\implies R(x) \qquad\qquad\qquad\qquad
		R(x) \implies E(x, 0) \\
		E(x, x') &\implies x' = -2 x + 9 \land R(x') \\
		&\qquad \lor (S(x', x'-1) \land E(x, x'-1))
		\lor (S(x', x'+1) \land E(x, x'+1)).
	\end{align}
	The program is non-terminating when $x = 3$.
	This is witnessed by a solution $\sigma$ for $(\clauses, \WFRs)$, which is given by $\sigma(R)(x) \coloneqq x = 3$, $\sigma(E)(x, x') \coloneqq x = 3 \land 0 \le x' \land x' \le 3$, and $\sigma(S)(x', x'') \coloneqq x'' = x' + 1 \land x'' \le 3$.
\end{myexample}

\section{CounterExample-Guided Inductive Synthesis (CEGIS)}
\label{sec:cegis}
We explain how CounterExample-Guided Inductive Synthesis~\cite{Solar-Lezama2006} (CEGIS for short) works for a given \PCSPWF{} $(\clauses,\WFRs)$ following~\cite{Unno2021}.
Then, we add the extraction of positive/negative examples to the CEGIS architecture, which enables our decision tree-based synthesizer to use a simplified form of examples.

CEGIS proceeds through the iterative interaction between a synthesizer and a validator (Fig.~\ref{fig:cegis}), in which they exchange examples and candidate solutions.

\begin{mydefinition}
	A formula $\phi$ is an \emph{example} of $\clauses$ if $\ftv{\phi} = \emptyset$ and $\bigwedge \clauses \models \phi$ hold.
	Given a set $\examples$ of examples of $\clauses$, a predicate substitution $\sigma$ is a \emph{candidate solution} for $(\clauses, \WFRs)$ that is consistent with $\examples$ if $\sigma$ is a solution for $(\examples,\WFRs)$.
\end{mydefinition}

\paragraph{Synthesizer.}
The input for a synthesizer is a set $\examples$ of examples of $\clauses$ collected from previous CEGIS iterations.
The synthesizer tries to find a candidate solution $\sigma$ consistent with $\examples$ instead of a genuine solution for $(\clauses, \WFRs)$.
If the candidate solution $\sigma$ is found, then $\sigma$ is passed to the validator.
If $\examples$ is unsatisfiable, then $\examples$ witnesses unsatisfiability of $(\clauses, \WFRs)$.
Details of our synthesizer is described in Sect.~\ref{sec:synth_rank}.

\paragraph{Validator.}
A validator checks whether the candidate solution $\sigma$ from the synthesizer is a genuine solution of $ (\clauses,\WFRs)$ by using SMT solvers.
That is, satisfiability of $\models \lnot \bigwedge \sigma(\clauses)$ is checked.
If $\models \lnot \bigwedge \sigma(\clauses)$ is not satisfiable, then $\sigma$ is a genuine solution of the original \PCSPWF{} $(\clauses,\WFRs)$, so the validator accepts this.
Otherwise, the validator adds new examples to the set $\examples$ of examples.
Finally, the synthesizer is invoked again with the updated set $\examples$ of examples.

If $\models \lnot \bigwedge \sigma(\clauses)$ is satisfiable, new examples are constructed as follows.
Using SMT solvers, the validator obtains an assignment $\theta$ to term variables such that $\models \lnot \theta(\psi)$ holds for some $\psi \in \sigma(\clauses)$.
By \eqref{eq:pcspwf_clause}, $\models \lnot \theta(\psi)$ is a clause of the form 
$\models \lnot \theta (\phi) \land \big( \bigwedge_{i=1}^{\ell} \lnot \sigma(X_i)(\theta(\seq{t}_i)) \big) \land \big( \bigwedge_{i=\ell+1}^m \sigma(X_i)(\theta(\seq{t}_i)) \big)$.
To prevent this counterexample from being found in the next CEGIS iteration again, the validator adds the following example to $\examples$.
\begin{equation}
	\bigvee_{i=1}^{\ell} X_i(\theta(\seq{t}_i)) \lor \bigvee_{i=\ell+1}^m \lnot X_i(\theta(\seq{t}_i)) \label{eq:example}
\end{equation}

The CEGIS architecture repeats this interaction between the synthesizer and the validator until a genuine solution for $(\clauses,\WFRs)$ is found or $\examples$ witnesses unsatisfiability of $(\clauses,\WFRs)$.

\paragraph{Extraction of positive/negative examples.}

Examples obtained in the above explanation are a bit complex to handle in our decision tree-based synthesizer: each example in $\examples$ is a disjunction~\eqref{eq:example} of literals, which may contain multiple predicate variables.

To simplify the form of examples, we extract from $\examples$ the sets $\examples_X^+$ and $\examples_X^-$ of \emph{positive examples} (i.e., examples of the form $X(\seq{v})$) and \emph{negative examples} (i.e., examples of the form $\lnot X(\seq{v})$) for each $X \in \fpv{\examples}$.
This allows us to synthesize a predicate $\sigma(X)$ for each predicate variable $X \in \fpv{\examples}$ separately.
For simplicity, we write $\seq{v} \in \examples_X^+$ and $\seq{v} \in \examples_X^-$ instead of $X(\seq{v}) \in \examples_X^+$ and $\lnot X(\seq{v}) \in \examples_X^-$.

The extraction is done as follows.
We first substitute for each predicate variable application $X(\seq{v})$ in $\examples$ a boolean variable $b_{X(\seq{v})}$ to obtain a SAT problem $\mathbf{SAT}(\examples)$.
Then, we use SAT solvers to obtain an assignment $\eta$ that is a solution for $\mathbf{SAT}(\examples)$.
If a solution $\eta$ exists, then we construct positive/negative examples from $\eta$; otherwise, $\examples$ is unsatisfiable.
\begin{mydefinition}
	Let $\eta$ be a solution for $\mathbf{SAT}(\examples)$.
	For each predicate variable $X \in \fpv{\examples}$, we define the set $\examples^{+}_{X}$ of \emph{positive examples} and the set $\examples^{+}_{X}$ of \emph{negative examples} under the assignment $\eta$ by $\examples^{+}_{X} \coloneqq \{ \seq{v} \mid \eta(b_{X(\seq{v})}) = \mathbf{true} \}$ and $\examples^{-}_{X} \coloneqq \{ \seq{v} \mid \eta(b_{X(\seq{v})}) = \mathbf{false} \}$.
\end{mydefinition}
Note that some of predicate variable applications $X(\seq{v})$ may not be assigned true nor false because they do not affect the evaluation of $\mathbf{SAT}(\examples)$.
Such predicate variable applications are discarded from $\{(\examples^{+}_{X}, \examples^{-}_{X})\}_{X \in \fpv{\examples}}$.

Our method uses the extraction of positive and negative examples when the validator passes examples to the synthesizer.
If $X \in \fpv{\examples} \cap \WFRs$, then we apply our ranking function synthesizer to $(\examples^{+}_{X}, \examples^{-}_{X})$.
If $X \in \fpv{\examples} \setminus \WFRs$, then we apply an invariant synthesizer.

We say a candidate solution $\sigma$ is consistent with $\{(\examples^{+}_{X}, \examples^{-}_{X})\}_{X \in \fpv{\examples}}$ if $\models \sigma(X)(\seq{v}^{+})$ and $\models \lnot \sigma(X)(\seq{v}^{-})$ hold for each predicate variable $X \in \fpv{\examples}$, $\seq{v}^{+} \in \examples^{+}_{X}$, and $\seq{v}^{-} \in \examples^{-}_{X}$.
If a candidate solution $\sigma$ is consistent with $\{(\examples^{+}_{X}, \examples^{-}_{X})\}_{X \in \fpv{\examples}}$, then $\sigma$ is also consistent with $\examples$.

Note that unsatisfiability of $\{(\examples^{+}_{X}, \examples^{-}_{X})\}_{X \in \fpv{\examples}}$ does not immediately implies unsatisfiability of $\examples$ nor $(\clauses, \WFRs)$ because $\{(\examples^{+}_{X}, \examples^{-}_{X})\}_{X \in \fpv{\examples}}$ depends on the choice of the assignment $\eta$.
Therefore, the CEGIS architecture need to be modified: if synthesizers find unsatisfiability of $\{(\examples^{+}_{X}, \examples^{-}_{X})\}_{X \in \fpv{\examples}}$, then we add the negation of an unsatisfiability core to $\examples$ to prevent using the same assignment $\eta$ again.

Note that some restricted forms of~\eqref{eq:example} have also been considered in previous work and are called implication examples in \cite{Garg2014} and implication/negation constraints in \cite{Champion2018}.
Our extraction of positive and negative examples is applicable to the general form of~\eqref{eq:example}.

\section{Ranking Function Synthesis}
\label{sec:synth_rank}
\newcommand{\CEGISExamples}{\examples^+}
\newcommand{\CEGISExamplesStartEnd}{\underline{\CEGISExamples}}
\newcommand{\ExampleSource}{\seq{v}}
\newcommand{\ExampleTarget}{\seq{v}'}
\newcommand{\LexGt}{\succ}

In this section, we describe one of the main contributions, that is, our decision tree-based synthesizer, which synthesizes a candidate well-founded relation $\sigma(R)$ from a finite set $\CEGISExamples_{R}$ of examples.
We assume that only positive examples are given because well-founded relations occur only positively in \PCSPWF{} for termination analysis (see Sect.~\ref{sec:problem}).
The aim of our synthesizer is to find a piecewise affine lexicographic ranking function $\seq{f}(\seq{x})$ for the given set $\CEGISExamples_R$ of examples.
Below, we fix a predicate variable $R \in \WFRs$ and omit the subscript $\CEGISExamples_R = \CEGISExamples$.

\subsection{Basic Definitions}
To represent piecewise affine lexicographic ranking functions, we use decision trees like the one in Figure~\ref{fig:example_decision_tree}.
Let $\seq{x} = (x_1, \dots, x_n)$ be the program variables where each $x_i$ ranges over $\mathbb{Z}$.
\begin{mydefinition}
	A \emph{decision tree} $D$ is defined by
	$D \coloneqq \seq{g}(\seq{x}) \mid \mathbf{if}\ h(\seq{x}) \ge 0\ \mathbf{then}\ D\allowbreak\ \mathbf{else}\ D$
	where $\seq{g}(\seq{x}) = (g_k(\seq{x}), \dots, g_0(\seq{x}))$ is a tuple of affine functions and $h(\seq{x})$ is an affine function.
	A \emph{segmentation tree} $S$ is defined as a decision tree with undefined leaves $\bot$: that is,
	$S \coloneqq \bot \mid \mathbf{if}\ h(\seq{x}) \ge 0\ \mathbf{then}\ S\ \mathbf{else}\ S$.
	For each decision tree D, we can canonically assign a segmentation tree by replacing the label of each leaf with $\bot$. This is denoted by $S(D)$.
	For each decision tree $D$, we denote the corresponding piecewise affine function by $\seq{f}_D(\seq{x}) : \mathbb{Z}^n \to \mathbb{Z}^{k+1}$.
\end{mydefinition}
Each leaf in a segmentation tree $S$ corresponds to a polyhedron.
We often identify the segmentation tree $S$ with the set of leaves of $S$ and a leaf with the polyhedron corresponding to the leaf.
For example, we say something like ``for each $L \in S$, $\seq{v} \in L$ is a point in the polyhedron $L$''.

Suppose we are given a segmentation tree $S$ and a set $\CEGISExamples$ of examples.
\begin{mydefinition}
	For each $L_1, L_2 \in S$, we denote the set of example transitions from $L_1$ to $L_2$ by $\CEGISExamples_{L_1, L_2} \coloneqq \{ (\ExampleSource, \ExampleTarget) \in \CEGISExamples \mid \ExampleSource \in L_1, \ExampleTarget \in L_2 \}$.
	An example $(\ExampleSource, \ExampleTarget) \in \CEGISExamples$ is \emph{crossing} w.r.t.\ $S$ if $(\ExampleSource, \ExampleTarget) \in \CEGISExamples_{L_1, L_2}$ for some $L_1 \neq L_2$, and \emph{non-crossing} if $(\ExampleSource, \ExampleTarget) \in \CEGISExamples_{L, L}$ for some $L$.
\end{mydefinition}
\begin{mydefinition}
	We define the \emph{dependency graph} $G(S, \CEGISExamples)$ for $S$ and $\CEGISExamples$ by the graph $(V, E)$ where vertices $V = S$ are leaves, and edges $E = \{ (L_1, L_2) \mid L_1 \neq L_2, \exists (\seq{v}, \seq{v}') \in \CEGISExamples_{L_1, L_2} \}$ are crossing examples.
\end{mydefinition}

We denote the set of start points $\seq{v}$ and end points $\seq{v}'$ of examples $(\seq{v}, \seq{v}') \in \CEGISExamples$ by $\CEGISExamplesStartEnd \coloneqq \{ \seq{v} \mid (\seq{v}, \seq{v}') \in \CEGISExamples \} \cup \{ \seq{v}' \mid (\seq{v}, \seq{v}') \in \CEGISExamples \}$.

\subsection{Segmentation and (Explicit and Implicit) Cycles: One-Dimensional Case}
For simplicity, we first consider the case where $\seq{f}(\seq{x}) = f(\seq{x}) : \mathbb{Z}^n \to \mathbb{Z}$ is a one-dimensional ranking function.
Our aim is to find a ranking function $f(\seq{x})$ for $\CEGISExamples$, which satisfies
$\forall (\ExampleSource, \ExampleTarget) \in \CEGISExamples.\ f(\ExampleSource) > f(\ExampleTarget)$
and
$\forall (\ExampleSource, \ExampleTarget) \in \CEGISExamples.\ f(\ExampleSource) \ge 0 \label{eq:ranking_condition}$.
If our ranking function synthesizer finds such a ranking function $f(\seq{x})$, then a candidate well-founded relation $R_f$ is constructed as
$R_f(\seq{x}, \seq{x}') \coloneqq f(\seq{x}) \ge 0 \land f(\seq{x}) > f(\seq{x}')$.

Our synthesizer builds a decision tree $D$ to find a ranking function $f_D(\seq{x})$ for $\CEGISExamples$.
The main question in doing so is ``when and how should we refine partitions of decision trees?''
To answer this question, we consider the case where there is no ranking function $f_D(\seq{x})$ for $\CEGISExamples$ with a fixed segmentation $S$, and classify reasons for this into three cases as follows.

\paragraph{Case 1: explicit cycles in examples.}
We define an \emph{explicit cycle} in $\CEGISExamples$ as a cycle in the graph $(\mathbb{Z}^n, \CEGISExamples)$.
An explicit cycle witnesses that there is no ranking function for $\CEGISExamples$ (see e.g., Example~\ref{ex:decision_tree_learning2}).

\paragraph{Case 2: non-crossing examples are unsatisfiable.}
The second case is when there is a leaf $L \in S$ such that no affine (not \emph{piecewise} affine) ranking function for the set $\CEGISExamples_{L, L}$ of non-crossing examples exists.
This prohibits the existence of piecewise affine function $f_D(\seq{x})$ for $\CEGISExamples$ with segmentation $S = S(D)$ because the restriction of $f_D(\seq{x})$ to $L \in S$ must be an affine ranking function for $\CEGISExamples_{L, L}$.

\paragraph{Case 3: implicit cycles in the dependency graph.}
We define an \emph{implicit cycle} by a cycle in the dependency graph $G(S, \CEGISExamples)$.
Case~3 is the case where an implicit cycle prohibits the existence of piecewise affine ranking functions for $\CEGISExamples$ with the segmentation $S$ (e.g., Example~\ref{ex:decision_tree_learning3}).
If Case~1 and Case~2 do not hold but no piecewise affine ranking function for $\CEGISExamples$ with the segmentation $S$ exists, then there must be an implicit cycle by (the contraposition of) the following proposition.
\begin{myproposition}\label{prop:sufficient_condition_decision_tree}
	Assume $\CEGISExamples$ is a set of examples that does not contain explicit cycles (i.e.\ Case~1 does not hold).
	Let $S$ be a segmentation tree and assume that for each $L \in S$, there exists an affine ranking function $f_L(\seq{x})$ for $\CEGISExamples_{L, L}$ (i.e.\ Case~2 does not hold).
	If the dependency graph $G(S, \CEGISExamples)$ is acyclic, then there exists a decision tree $D$ with the segmentation $S(D) = S$ such that $f_{D}(\seq{x})$ is a ranking function for $\CEGISExamples$.
\end{myproposition}
\begin{proof}
	By induction on the height (i.e.\ the length of a longest path from a vertex) of vertices in $G(S, \CEGISExamples)$.
	We construct a decision tree $D$ as follows.
	If the height of $L \in S$ is 0, then we assign $f'_L(\seq{x}) \coloneqq f_L(\seq{x})$ to the leaf $L$ where $f_L(\seq{x})$ is a ranking function for $\CEGISExamples_{L, L}$.
	If the height of $L \in S$ is $n > 0$, then we assign $f'_{L}(\seq{x}) \coloneqq f_L(\seq{x}) + c$ to the leaf $L$ where $c \in \mathbb{Z}$ is a constant that satisfies $\forall (\ExampleSource, \ExampleTarget) \in \CEGISExamples_{L, L'}, f_{L}(\ExampleSource) + c > f'_{L'}(\ExampleTarget)$ for each cell $L'$ with the height less than $n$.
	\qed
\end{proof}
Note that the converse of Proposition~\ref{prop:sufficient_condition_decision_tree} does not hold: the existence of implicit cycles in $G(S, \CEGISExamples)$ does not necessarily imply that no piecewise affine ranking function exists with the segmentation $S$.

\subsection{Segmentation and (Explicit and Implicit) Cycles: Multi-Dimensional Lexicographic Case}
\label{subsec:synth_lexicographic_ranking}
We consider a more general case where $\seq{f}(\seq{x}) = (f_k(\seq{x}), \dots, f_0(\seq{x}))$ is a multi-dimensional lexicographic ranking function and $k$ is a fixed nonnegative integer.

Given a function $\seq{f}(\seq{x})$, we consider the well-founded relation $R_{\seq{f}}(\seq{x}, \seq{x}')$ defined inductively as follows.
\begin{equation}
	R_{()}(\seq{x}, \seq{x}') \coloneqq \bot \quad
	\begin{array}[t]{ll}
		R_{(f_{k}, \dots, f_0)}(\seq{x}, \seq{x}') &\coloneqq f_{k}(\seq{x}) \ge 0 \land f_{k}(\seq{x}) > f_{k}(\seq{x}') \\
		&\quad\lor f_{k}(\seq{x}) = f_{k}(\seq{x}') \land R_{(f_{k-1}, \dots, f_0)}(\seq{x}, \seq{x}')
	\end{array} \label{eq:well_founded_lexicographic1}
\end{equation}

Our aim here is to find a lexicographic ranking function $\seq{f}(\seq{x})$ for $\CEGISExamples$, i.e.\ a function $\seq{f}(\seq{x})$ such that $R_{\seq{f}}(\ExampleSource, \ExampleTarget)$ holds for each $(\ExampleSource, \ExampleTarget) \in \CEGISExamples$.
Our synthesizer does so by building a decision tree.
The same argument as the one-dimensional case holds for lexicographic ranking functions.
\begin{mytheorem}[cycle detection]\label{thm:sufficient_condition_decision_tree_lexicographic}
	Assume $\CEGISExamples$ is a set of examples that does not contain explicit cycles.
	Let $S$ be a segmentation tree and assume that for each $L \in S$, there exists an affine function $\seq{f}_L(\seq{x})$ that satisfies $\forall (\ExampleSource, \ExampleTarget) \in \CEGISExamples_{L, L}, R_{\seq{f}_L}(\ExampleSource, \ExampleTarget)$.
	If the dependency graph $G(S, \CEGISExamples)$ is acyclic, then there exists a decision tree $D$ with the segmentation $S(D) = S$ such that $R_{\seq{f}_{D}}(\ExampleSource, \ExampleTarget)$ holds for each $(\ExampleSource, \ExampleTarget) \in \CEGISExamples$.
\end{mytheorem}
\begin{proof}
	The proof is almost the same as Proposition~\ref{prop:sufficient_condition_decision_tree}.
	Here, note that if $\seq{f}'(\seq{x}) = \seq{f}(\seq{x}) + \seq{c}$ where $\seq{c}$ is a tuple of nonnegative integer constants, then $R_{\seq{f}'}(\seq{x}, \seq{x}')$ subsumes $R_{\seq{f}}(\seq{x}, \seq{x}')$.
	\qed
\end{proof}

\subsection{Our Decision Tree Learning Algorithm}\label{subsec:dt_alg}
We design a concrete algorithm based on Theorem~\ref{thm:sufficient_condition_decision_tree_lexicographic}.
It is shown in Algorithm~\ref{alg:build_decision_tree} and consists of three phases.  We shall describe the three phases one by one.

\begin{algorithm}[tbp]
	\caption{Building decision trees.}
	\label{alg:build_decision_tree}
	\begin{algorithmic}[1]
		\Require{a set $\CEGISExamples$ of examples, an integer $k \ge 0$}
		\Ensure{a well-founded relation $R$ such that $\forall (\seq{x}, \seq{x}') \in \CEGISExamples, R(\seq{x}, \seq{x}')$}
		\If{$E$ has a cycle} \label{alg:line:cycle_detection_begin}
			\State\Return unsatisfiable
		\EndIf \label{alg:line:cycle_detection_end}
		\State $D \coloneqq \Call{ResolveCase2}{E}$ \label{alg:line:call_resolve_case2}
		\While{true} \label{alg:line:resolve_case3_begin}
			\State $C \coloneqq \Call{GetConstraints}{D, E}$ \label{alg:line:build_dt_get_constraints}
			\State $O \coloneqq \Call{SumAbsParams}{D}$
			\State $\rho \coloneqq \Call{Minimize}{O, C}$ \label{alg:line:minimize}
			\If{$\rho$ is defined}
				\State $\seq{f}(\seq{x}) \coloneqq \seq{f}_{\rho(D)}(\seq{x})$
				\State\Return $R_{\seq{f}}$ \label{alg:line:return_candidate}
			\Else
				\State get an unsat core in $C$
				\State find an implicit cycle $(\ExampleSource_1, \ExampleTarget_1), \dots, (\ExampleSource_l, \ExampleTarget_l)$ in the unsat core \label{alg:line:find_spurious_cycle}
				\State find a cell $C$ and two distinct points $\ExampleTarget_i, \ExampleSource_{i+1} \in C$ in the implicit cycle
				\State add a halfspace to separate $\ExampleTarget_i$ and $\ExampleSource_{i+1}$ and update $D$ \label{alg:line:update_dt_case3}
			\EndIf
		\EndWhile \label{alg:line:resolve_case3_end}
	\end{algorithmic}
\end{algorithm}
\begin{algorithm}[tb]
	\caption{Resolving Case~2.}
	\label{alg:resolve_case2}
	\begin{algorithmic}[1]
		\Function{ResolveCase2}{$\examples'^{+}$}
			\State $\seq{f} \coloneqq \Call{MakeAffineTemplate}{k}$ \label{alg:resolve_case2:make_template}
			\State $C \coloneqq \Call{GetConstraints}{\seq{f}, \examples'^{+}}$ \label{alg:line:resolve_case2_get_constraints}
			\State $\rho \coloneqq \Call{GetModel}{C}$
			\If{$\rho$ is undefined}
				\State $h \coloneqq \Call{ChooseQualifier}{\examples'^{+}}$ \label{alg:line:choose_halfspace}
				\State $D_{\ge 0} \coloneqq \Call{ResolveCase2}{\{ (\seq{v}, \seq{v}') \in \examples'^{+} \mid h(\seq{v}) \ge 0 \land h(\seq{v}') \ge 0 \}}$
				\State $D_{< 0} \coloneqq \Call{ResolveCase2}{\{ (\seq{v}, \seq{v}') \in \examples'^{+} \mid h(\seq{v}) < 0 \land h(\seq{v}') < 0 \}}$
				\State\Return $(\mathbf{if}\ h(\seq{x}) \ge 0\ \mathbf{then}\ D_{\ge 0}\ \mathbf{else}\ D_{< 0})$
			\Else
				\State\Return $\seq{f}$
			\EndIf
		\EndFunction
		\Function{GetConstraints}{$D, \CEGISExamples$}
			\State\Return $\{ R_{\seq{f}_D}(\seq{v}, \seq{v}') \mid (\seq{v}, \seq{v}') \in \CEGISExamples \}$ where $\seq{f}_D$ is the tuple of piecewise affine functions corresponding to $D$
		\EndFunction
	\end{algorithmic}
\end{algorithm}

\vspace{-1em}
\begin{algorithm}[tb]
	\caption{A criterion for eager qualifier selection.}
	\label{alg:rank_qualifier}
	\begin{algorithmic}[1]
		\Function{QualityMeasure}{$h, \examples'^{+}$}
			\State $E_{++} \coloneqq \{ (\seq{v}, \seq{v}') \in \examples'^{+} \mid h(\seq{v}) \ge 0 \land h(\seq{v}) \ge 0 \}$
			\State $E_{+-} \coloneqq \{ (\seq{v}, \seq{v}') \in \examples'^{+} \mid h(\seq{v}) \ge 0 \land h(\seq{v}) < 0 \}$
			\State $E_{-+} \coloneqq \{ (\seq{v}, \seq{v}') \in \examples'^{+} \mid h(\seq{v}) < 0 \land h(\seq{v}) \ge 0 \}$
			\State $E_{--} \coloneqq \{ (\seq{v}, \seq{v}') \in \examples'^{+} \mid h(\seq{v}) < 0 \land h(\seq{v}) < 0 \}$
			\State $\seq{f} \coloneqq \Call{MakeAffineTemplate}{k}$
			\State \makebox[17em]{$C_{+} \coloneqq \Call{GetConstraints}{\seq{f}, E_{++}}$\hfill} $C_{-} \coloneqq \Call{GetConstraints}{\seq{f}, E_{--}}$
			\State \makebox[17em]{$N_{+} \coloneqq \Call{MaxSmt}{C_{+}}$\hfill} $N_{-} \coloneqq \Call{MaxSmt}{C_{-}}$
			\State\Return $N_{+} + N_{-} + (|E_{+-}| + |E_{-+}|) (1 - \mathrm{entropy}(|E_{+-}|, |E_{-+}|))$
		\EndFunction
	\end{algorithmic}
\end{algorithm}

\subsubsection{Phase~1}
Phase~1 (Line~\ref{alg:line:cycle_detection_begin}-\ref{alg:line:cycle_detection_end}) detects explicit cycles in $\CEGISExamples$ to exclude Case~1.
Here, we use a cycle detection algorithm for directed graphs.

\vspace{-1em}
\subsubsection{Phase~2}
Phase~2 (Line~\ref{alg:line:call_resolve_case2}) detects and resolves Case~2 by using \textsc{ResolveCase2} (Algorithm~\ref{alg:resolve_case2}), which is a function that grows a decision tree recursively.
\textsc{ResolveCase2} takes non-crossing examples in a leaf, divides the leaf, and returns a \emph{template tree} that is fine enough to avoid Case~2.
Here, template trees are decision trees whose leaves are labeled by affine templates.

Algorithm~\ref{alg:resolve_case2} shows the detail of \textsc{ResolveCase2}.
\textsc{ResolveCase2} builds a template tree recursively starting from the trivial segmentation $S = \bot$ and all given examples.
In each polyhedron, \textsc{ResolveCase2} checks whether the set $C$ of constraints imposed by non-crossing examples can be satisfied by an affine lexicographic ranking function on the polyhedron (Line~\ref{alg:resolve_case2:make_template}-\ref{alg:line:resolve_case2_get_constraints}).
If the set $C$ of constraints is not satisfiable, then \textsc{ResolveCase2} chooses a halfspace $h(\seq{x}) \ge 0$ (Line~\ref{alg:line:choose_halfspace}) and divides the current polyhedron by the halfspace.

There is a certain amount of freedom in the choice of halfspaces.
To guarantee termination of the whole algorithm, we require that the chosen halfspace $h$ separates at least one point in $\underline{\examples'^{+}} \coloneqq \{ \seq{v} \mid (\seq{v}, \seq{v}') \in \examples'^{+} \} \cup \{ \seq{v}' \mid (\seq{v}, \seq{v}') \in \examples'^{+} \}$ from the other points in $\underline{\examples'^{+}}$.
That is:
\begin{myassumption}\label{assump:qualifier}
	If halfspace $h(\seq{x}) \ge 0$ is chosen in Line~\ref{alg:line:choose_halfspace} of Algorithm~\ref{alg:resolve_case2}, then there exist $\seq{v}, \seq{u} \in \underline{\examples'^{+}}$ such that $h(\seq{v}) \ge 0$ and $h(\seq{u}) < 0$.
\end{myassumption}

We explain two strategies (eager and lazy) to choose halfspaces that can be used to implement \textsc{ChooseQualifier}.
Both of them are guaranteed to terminate, and moreover, intended to yield simple decision trees.

\paragraph{Eager strategy.}
In the eager strategy, we eagerly generate a finite set $H$ of halfspaces from the set $\CEGISExamples$ of all examples beforehand and choose the best one from $H$ with respect to a certain quality measure.
To satisfy Assumption~\ref{assump:qualifier}, $H$ are generated so that any two points $\seq{u}, \seq{v} \in \CEGISExamplesStartEnd$ can be separated by some halfspace $(h(\seq{x}) \ge 0) \in H$.

For example, we can use intervals $H = \{\pm (x_i - a_i) \ge 0 \mid i = 1, \dots, n \land (a_1, \dots, a_n) \in \underline{\CEGISExamples} \}$ and octagons $H = \{\pm (x_i - a_i) \pm (x_j - a_j) \ge 0 \mid i \neq j \land (a_1, \dots, a_n) \in \CEGISExamplesStartEnd \}$ where $\seq{x} = (x_1, \dots, x_n)$.
For any input $\examples'^{+} \subseteq \CEGISExamples$ of \textsc{ResolveCase2}, intervals and octagons satisfy $\emptyset \neq H' \coloneqq \{ h(\seq{x}) \ge 0 \mid \exists \seq{v}, \seq{u} \in \examples'^{+}. h(\seq{v}) \ge 0 \land h(\seq{u}) < 0 \}$, so Assumption~\ref{assump:qualifier} is satisfied by choosing the best halfspace with respect to the quality measure from $H'$.

For each halfspace $(h(\seq{x}) \ge 0) \in H'$, we calculate \textsc{QualityMeasure} in Algorithm~\ref{alg:rank_qualifier}, and choose one that maximizes \Call{QualityMeasure}{$h, \examples'^{+}$}.
\Call{QualityMeasure}{$h, \examples'^{+}$} calculates the sum of the maximum number of satisfiable constraints in each leaf divided by $h(\seq{x}) \ge 0$ plus an additional term $(|E_{+-}| + |E_{-+}|) (1 - \mathrm{entropy}(|E_{+-}|, |E_{-+}|))$ where
$\mathrm{entropy}(x, y) = - \frac{x}{x + y} \log_2 \frac{x}{x + y} - \frac{y}{x + y} \allowbreak \log_2 \frac{y}{x + y}$.
Therefore, the term $(|E_{+-}| + |E_{-+}|) (1 - \mathrm{entropy}(|E_{+-}|, |E_{-+}|))$ is close to $|E_{+-}| + |E_{-+}|$ if almost all examples in $E_{+-} \cup E_{-+}$ cross $h$ in the same direction and close to $0$ if $|E_{+-}|$ is almost equal to $|E_{-+}|$.

\paragraph{Lazy strategy.}
In the lazy strategy, we lazily generate halfspaces.
We divide the current polyhedron so that non-crossing examples in the cell point to almost the same direction.

First, we label states that occur in $\CEGISExamples_{C, C}$ as follows.
We find a direction that most examples in $C$ point to by solving the MAX-SMT
$\vec{a} \coloneqq \max_{\vec{a}} \big|\{ (\ExampleSource, \ExampleTarget) \in \CEGISExamples_{C, C} \mid \vec{a} \cdot (\ExampleSource - \ExampleTarget) > 0 \} \big|$.
For each $(\ExampleSource, \ExampleTarget) \in \CEGISExamples_{C, C}$, we label two points $\ExampleSource, \ExampleTarget$ with $+1$ if $\vec{a} \cdot (\ExampleSource - \ExampleTarget) > 0$ and with $-1$ otherwise.

Then we apply weighted C-SVM to generate a hyperplane that separates most of the positive and negative points.
To guarantee termination of Algorithm~\ref{alg:build_decision_tree}, we avoid ``useless'' hyperplanes that classify all the points by the same label.
If we obtain such a useless hyperplane, then we undersample a majority class and apply C-SVM again.
By undersampling suitably, we eventually get linearly separable data with at least one positive point and one negative point.

Note that since coefficients of hyperplanes extracted from C-SVM are floating point numbers, we have to approximate them by hyperplanes with rational coefficients.
This is done by truncating continued fraction expansions of coefficients by a suitable length.

\vspace{-1em}
\subsubsection{Phase~3}
In Line~\ref{alg:line:resolve_case3_begin}-\ref{alg:line:resolve_case3_end} of Algorithm~\ref{alg:build_decision_tree}, we further refine the segmentation $S(D)$ to resolve Case~3.
 Once Case~2 is resolved by \textsc{ResolveCase2}, Case~2 never holds even after refining $S(D)$ further. This enables to separate Phases~2 and~3.

Given a template tree $D$, we consider the set $C$ of constraints on parameters in $D$ that claims $\seq{f}_D(\seq{x})$ is a ranking function for $\CEGISExamples$ (Line~\ref{alg:line:build_dt_get_constraints}).

If $C$ is satisfiable, we use an SMT solver to obtain a solution of $C$ (i.e.\ an assignment $\rho$ of integers to parameters) while minimizing the sum of absolute values of unknown parameters in $D$ at the same time (Line~\ref{alg:line:minimize}).
This minimization is intended to give a simple candidate ranking function.
The solution $\rho$ is used to instantiate the template tree $D$ (Line~\ref{alg:line:return_candidate}).

If $C$ cannot be satisfied, there must be an implicit cycle in the dependency graph $G(S(D), \CEGISExamples)$ by Theorem~\ref{thm:sufficient_condition_decision_tree_lexicographic}.
The implicit cycle can be found in an unsatisfiable core of $C$.
We refine the segmentation of $D$ to cut the implicit cycle in Line~\ref{alg:line:update_dt_case3}.
To guarantee termination, we choose a halfspace satisfying the following assumption, which is similar to Assumption~\ref{assump:qualifier}.
\begin{myassumption}\label{assump:qualifier2}
	If halfspace $h(\seq{x}) \ge 0$ is chosen in Line~\ref{alg:line:update_dt_case3} of Algorithm~\ref{alg:build_decision_tree}, then there exist $\seq{v}, \seq{u} \in \CEGISExamplesStartEnd$ such that $h(\seq{v}) \ge 0$ and $h(\seq{u}) < 0$.
\end{myassumption}

We have two strategy (eager and lazy) to refine the segmentation of $D$.

In eager strategy, we choose a halfspace $(h(\seq{x}) \ge 0) \in H$ that separates two distinct points $\ExampleTarget_i$ and $\ExampleSource_{i+1}$ in the implicit cycle.
In doing so, we want to reduce the number of implicit cycles in $G(S(D), \CEGISExamples)$, but adding a new halfspace may introduce new implicit cycles if there exists $(\ExampleSource, \ExampleTarget) \in \CEGISExamples_{C, C}$ that crosses the new border from the side of $\ExampleTarget_i$ to the side of $\ExampleSource_{i+1}$.
Therefore, we choose a hyperplane that minimizes the number of new crossing examples.

In lazy strategy, we use an SMT solver to find a hyperplane $h(\seq{x}) \in H$ that separates $\ExampleTarget_i$ and $\ExampleSource_{i+1}$ and minimizes the number of new crossing examples.

\subsubsection{Termination}
Assumption~\ref{assump:qualifier} and Assumption~\ref{assump:qualifier2} guarantees that every leaf in $S(D)$ contains at least one point in the finite set $\CEGISExamplesStartEnd$.
Because the number of leaves in $S(D)$ strictly increases after each iteration of Phase~2 and Phase~3, we eventually get a segmentation $S(D)$ where each $L \in S(D)$ contains only one point in $\CEGISExamplesStartEnd$ in the worst case.
Since we have excluded Case~1 at the beginning, Theorem~\ref{thm:sufficient_condition_decision_tree_lexicographic} guarantees the existence of ranking function with the segmentation $S(D)$.
Therefore, the algorithm terminates within $|\CEGISExamplesStartEnd|$ times of refinement.

\begin{mytheorem}
	If Assumption~\ref{assump:qualifier} and Assumption~\ref{assump:qualifier2} hold, then Algorithm~\ref{alg:build_decision_tree} terminates.
	If Algorithm~\ref{alg:build_decision_tree} returns a piecewise affine lexicographic function $\seq{f}(\seq{x})$, then the function satisfies $R_{\seq{f}}(\seq{x}, \seq{x}')$ for each $(\seq{x}, \seq{x}') \in \CEGISExamples$ where $\CEGISExamples$ is the input of the algorithm.
	\qed
\end{mytheorem}

\subsection{Improvement by Degenerating Negative Values}\label{subsec:better_well_founded_relation}
There is another way to define well-founded relation from the tuple $\seq{f}(\seq{x}) = (f_k(\seq{x}), \dots, f_0(\seq{x}))$ of functions, that is, the well-founded relation $R'_{\seq{f}}(\seq{x}, \seq{x}')$ defined inductively by $R'_{()}(\seq{x}, \seq{x}') \coloneqq \bot$ and $R'_{(f_{k}, \dots, f_0)}(\seq{x}, \seq{x}') \coloneqq f_{k}(\seq{x}) \ge 0 \land f_{k}(\seq{x}) > f_{k}(\seq{x}') \lor \big(f_{k}(\seq{x}') < 0 \lor f_{k}(\seq{x}) = f_{k}(\seq{x}') \big) \land R'_{(f_{k-1}, \dots, f_0)}(\seq{x}, \seq{x}')$.

In this definition, we loosen the equality $f_{i}(\seq{x}) = f_{i}(\seq{x}')$ (where $i = 1, \dots, k$) of the usual lexicographic ordering \eqref{eq:well_founded_lexicographic1} to $f_{i}(\seq{x}') < 0 \lor f_{i}(\seq{x}) = f_{i}(\seq{x}')$.
This means that once $f_{i}(\seq{x})$ becomes negative, $f_i(\seq{x})$ must stay negative but the value do not have to be the same, which is useful for the synthesizer to avoid complex candidate lexicographic ranking functions and thus improves the performance.

However, if we use this well-founded relation $R'_{\seq{f}}(\seq{x}, \seq{x}')$ instead of $R_{\seq{f}}(\seq{x}, \seq{x}')$ in \eqref{eq:well_founded_lexicographic1}, then Theorem~\ref{thm:sufficient_condition_decision_tree_lexicographic} fails because $R'_{\seq{f}}(\seq{x}, \seq{x}')$ is not necessarily subsumed by $R'_{\seq{f} + \seq{c}}$ where $\seq{c} = (c_k, \dots, c_0)$ is a nonnegative constant (see the proof of Proposition~\ref{prop:sufficient_condition_decision_tree} and Theorem~\ref{thm:sufficient_condition_decision_tree_lexicographic}).
As a result, there is a chance that no implicit cycle can be found in line~\ref{alg:line:find_spurious_cycle} of Algorithm~\ref{alg:build_decision_tree}.
Therefore, when we use $R'_{\seq{f}}(\seq{x}, \seq{x}')$, we modify Algorithm~\ref{alg:build_decision_tree} so that if no implicit cycle can be found in line~\ref{alg:line:find_spurious_cycle}, then we fall back on the former definition of $R_{\seq{f}}(\seq{x}, \seq{x}')$ and restart Algorithm~\ref{alg:build_decision_tree}.

\section{Implementation and Evaluation}
\label{sec:eval}
\paragraph{Implementation.}
We implemented a constraint solver \muval{} that supports invariant synthesis and ranking function synthesis.
For invariant synthesis, we apply an ordinary decision tree learning (see~\cite{Krishna2015,Garg2016,Champion2018,Ezudheen2018,Zhu2018} for existing techniques).
For ranking function synthesis, we implemented the algorithm in Sect.~\ref{sec:synth_rank} with both eager and lazy strategies for halfspace selection.
Our synthesizer uses well-founded relation explained in Sect.~\ref{subsec:better_well_founded_relation}.
Given a benchmark, we run our solver for both termination and non-termination verification in parallel, and when one of the two returns an answer, we stop the other and use the answer.
\muval{} is written in OCaml and uses Z3 as an SMT solver backend.
We used clang and llvm2kittel~\cite{llvm2kittel} to convert C benchmarks to T2~\cite{t2github} format files, which are then translated to \PCSPWF{} by \muval{}.

\paragraph{Experiments.}
We evaluated our implementation \muval{} on C benchmarks from Termination Competition 2020 (C Integer)~\cite{termcomp2020}.
We compared our tool with \aprove{}~\cite{Emmes2012, Brockschmidt2012}, \irankfinder{}~\cite{BenAmram2014}, and \ultimateautomizer{}~\cite{Heizmann2014}.
Experiments are conducted on StarExec~\cite{starexec} (CentOS 7.7 (1908) on Intel(R) Xeon(R) CPU E5-2609 0 @ 2.40GHz (2393 MHZ) with 263932744 kB main memory).
The time limit was 300 seconds.

\begin{wraptable}[8]{r}{18em}
	\centering
	\vspace*{-2.5em}
	\caption{Numbers of solved benchmarks}
	\label{tab:result}
	\vspace*{-0.5em}
	\begin{tabular}{c|c@{\hspace{1ex}}c@{\hspace{1ex}}c@{\hspace{1ex}}c}
		& Yes & No & TO & U \\
		\hline
		\muval{} (eager) & 204 & 89 & 42 & 0 \\
		\muval{} (lazy) & 200 & 84 & 51 & 0\\
		\aprove{} & 216 & 100 & 16 & 3 \\
		\irankfinder{} & 208 & 92\footnotemark & 0 & 34 \\
		\ultimateautomizer{} & 180 & 83 & 2 & 70
	\end{tabular}
\end{wraptable}
\footnotetext{We removed one benchmarks from the result of \irankfinder{} because the answer was wrong.}

\begin{figure}[tbp]
	\centering
	\includegraphics[width=0.3\linewidth,trim=40 0 40 0]{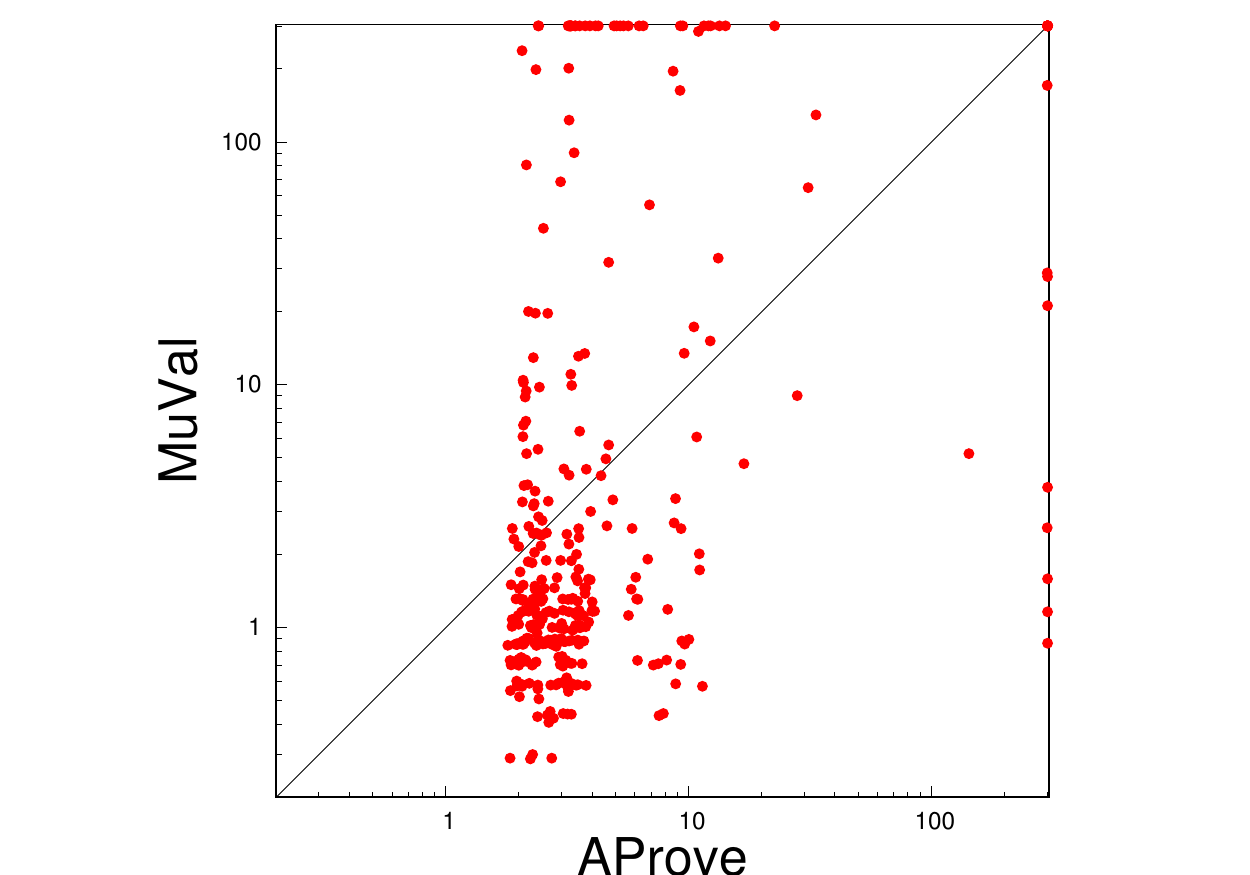}
	\includegraphics[width=0.3\linewidth,trim=40 0 40 0]{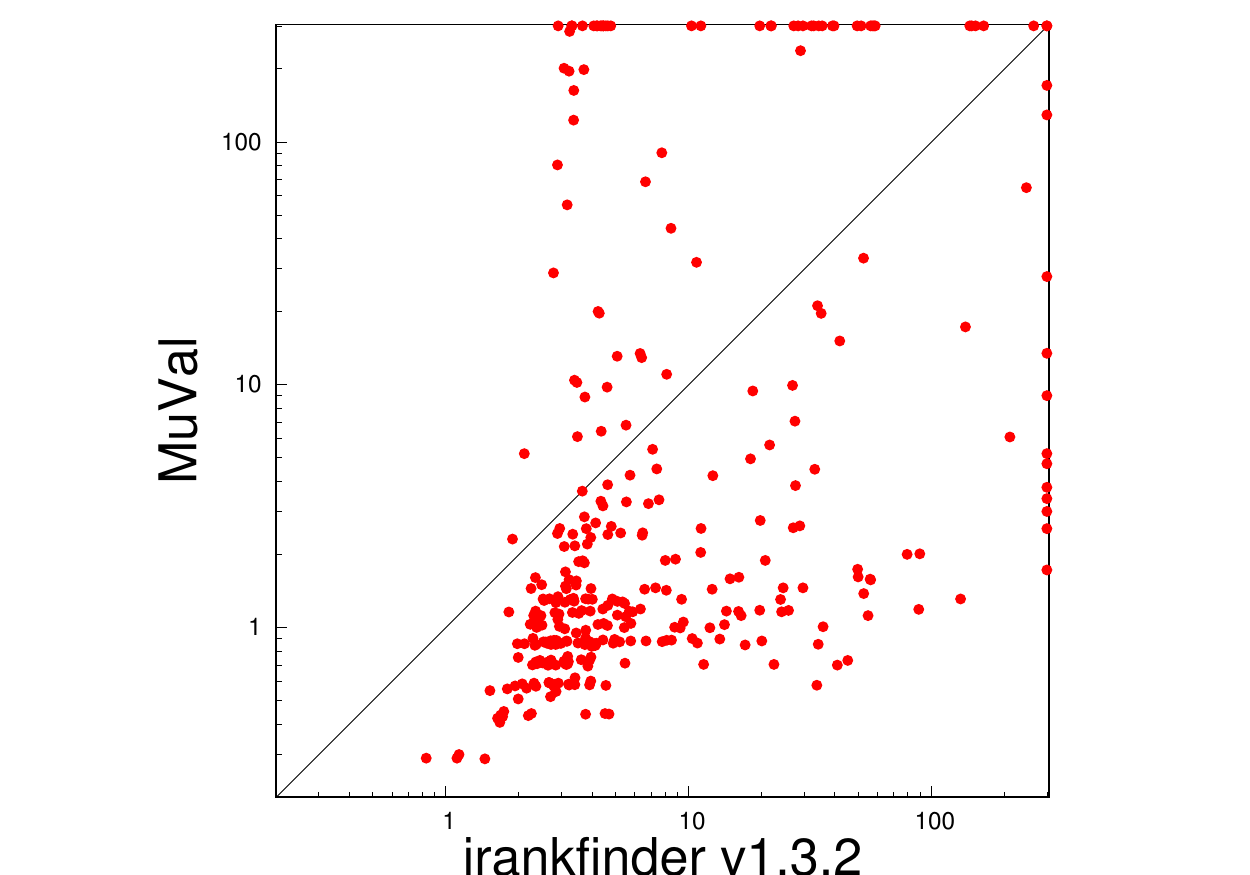}
	\includegraphics[width=0.3\linewidth,trim=40 0 40 0]{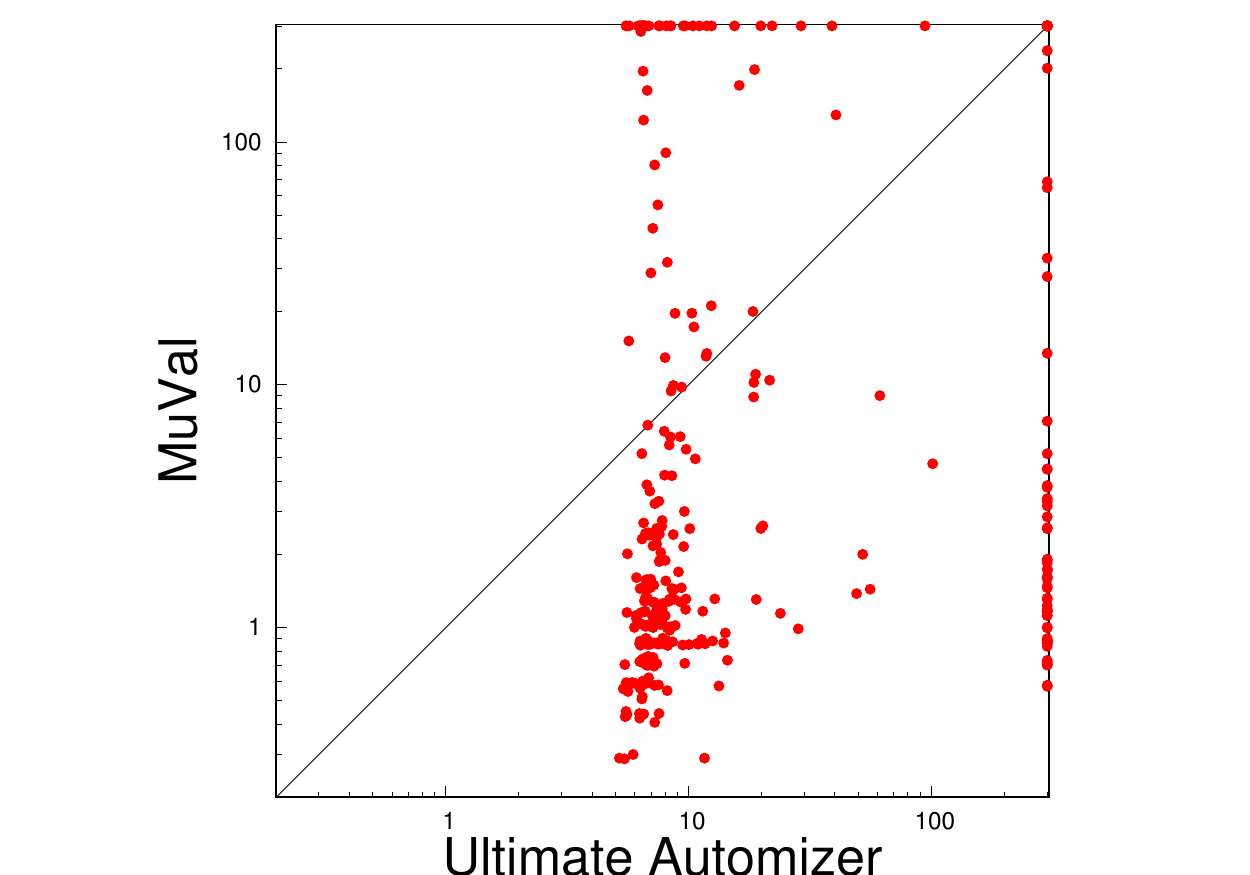}
	\caption{Scatter plots of runtime. \ultimateautomizer{} and \aprove{} sometimes gave up before the time limit, and such cases are regarded as 300s.}
	\label{fig:result_scatter}
\end{figure}

\paragraph{Results.}
Results are shown in Table~\ref{tab:result}.
Yes/No/TO/U means the number of benchmarks that these tools could verify termination/could verify non-termination/could not answer within 300 seconds and timed out (\underline{T}ime\underline{O}ut)/gave up before 300 seconds (\underline{U}nknown), respectively.
We also show scatter plots of runtime in Fig.~\ref{fig:result_scatter}.

\muval{} was able to solve more benchmarks than \ultimateautomizer{}.
Compared to \irankfinder{}, \muval{} solved slightly fewer benchmarks, but was faster in a large number of benchmarks: 265 benchmarks were solved faster by \muval{}, 68 by \irankfinder{}, and 2 were not solved by both tools within 300 seconds (here, we regard U (unknown) as 300 seconds).
Compared to \aprove{}, \muval{} solved fewer benchmarks.
However, there are several benchmarks that \muval{} could solve but \aprove{} could not.
Among them is ``TelAviv-Amir-Minimum\_true-termination.c'', which does require piecewise affine ranking functions.
\muval{} found a ranking function $f(x, y) = \mathbf{if}\ x - y \ge 0\ \mathbf{then}\ y\ \mathbf{else}\ x$, while \aprove{} timed out.

We also observed that using CEGIS with transition examples itself showed its strengths even for benchmarks that do not require piecewise affine ranking functions.
Notably, there are three benchmarks that \muval{} could solve but the other tools could not; they are examples that do not require segmentations.
Further  analysis of these benchmarks indicates the following strengths of our framework:
 (1) the ability to handle nonlinear constraints (to some extent) thanks to the example-based synthesis and the recent development of SMT solvers; and (2) the ability to find a long lasso-shaped non-terminating trace assembled from multiple transition examples.
See \referappendix{appx:detail_experiment}{A} for details.

\section{Related Work}
\label{sec:related}
There are a bunch of works that synthesize ranking functions via constraint solving.
Among them is a counterexample-guided method like CEGIS~\cite{Solar-Lezama2006}.
CEGIS is sound but not guaranteed to be complete in general: even if a given constraint has a solution, CEGIS may fail to find the solution.
A complete method for ranking function synthesis is proposed in~\cite{Gonnord2015}.
They collect only extremal counterexamples instead of arbitrary transition examples to avoid infinitely many examples.
A limitation of their method is that the search space is limited to (lexicographic) affine ranking functions.

Another counterexample-guided method is proposed in~\cite{Urban2016} and implemented in \seahorn{}.
This method can synthesize piecewise affine functions, but their approach is quite different from ours.
Given a program, they construct a \emph{safety} property that the number of loop iterations does not exceed the value of a candidate ranking function.
The safety property is checked by a verifier.
If it is violated, then a trace is obtained as a counterexample and the candidate ranking function is updated by the counterexample.
The main difference from our method is that their method uses trace examples while our method uses transition examples (which is less expensive to handle).
\freqterm{}~\cite{Fedyukovich2018} also uses the connection to safety property, but they exploit syntax-guided synthesis for synthesizing ranking functions.

Aside from counterexample-guided methods, constraint solving is widely studied for affine ranking functions~\cite{Podelski2004}, lexicographic affine ranking functions~\cite{Alias2010,BenAmram2014,Leike2014a}, and multiphase affine ranking functions~\cite{Ben-Amram2017,BenAmram2019}. Their implementation includes \rankfinder\ and \irankfinder.
Farkas' lemma or Motzkin's transposition theorem are often used as a tool to transform $\exists\forall$-constraints to $\exists$-constraints.
However, when we apply this technique to piecewise affine ranking functions, we get nonlinear constraints~\cite{Leike2014a}.

Abstract interpretation is also applied to segmented synthesis of ranking functions and implemented in \function{}~\cite{Urban2013,Urban2014,Urban2014a}.
In this series of work, decision tree representation of ranking functions is used in~\cite{Urban2014a} for better handling of disjunctions.
Compared to their work, we believe that our method is more easily extensible to other theories than linear integer arithmetic as long as the theories are supported by SMT solvers (although such extensions are out of the scope of this paper).

Other state-of-the-art termination verifiers include the following.
\ultimateautomizer{}~\cite{Heizmann2014} is an automata-based method.
It repeatedly finds a trace and computes a termination argument that contains the trace until termination arguments cover the set of all traces.
B\"uchi automata are used to handle such traces.
\aprove{}~\cite{Emmes2012, Brockschmidt2012} is based on term rewriting systems.

\section{Conclusions and Future Work}
\label{sec:conc}
In this paper, we proposed a novel decision tree-based synthesizer for ranking functions, which is integrated into the CEGIS architecture.
The key observation here was that we need to cope with explicit and implicit cycles contained in given examples.
We designed a decision tree learning algorithm using the theoretical observation of the cycle detection theorem.
We implemented the framework and observed that its performance is comparable to state-of-the-art termination analyzers. In particular, it solved three benchmarks that no other tool solved, a result that demonstrates the potential of the current combination of CEGIS, segmented synthesis, and transition examples.

We plan to extend our ranking function synthesizer to a synthesizer of piecewise affine ranking supermartingales. Ranking supermartingales~\cite{chakarov2013} are probabilistic version of ranking functions and used for verification of almost-sure termination of probabilistic programs.

We also plan to implement a mechanism to automatically select a suitable set of halfspaces with which decision trees are built. In our ranking function synthesizer, intervals/octagons/octahedron/polyhedra can be used as the set of halfspaces. However, selecting an overly expressive set of halfspaces may cause the problem of overfitting~\cite{Padhi2019} and result in poor performance. Therefore, applying heuristics that adjusts the expressiveness of halfspaces based on the current examples may improve the performance of our tool.

\paragraph{Acknowledgement.}
We thank Andrea Peruffo and the anonymous referees for many suggestions.
This work was supported by JST ERATO HASUO Metamathematics for Systems Design Project (No. JPMJER1603) and JSPS KAKENHI Grant Numbers 20H04162, 20H05703, 19H04084, and 17H01720.

\bibliographystyle{splncs04}
\bibliography{abbrv,prog_lang}

\ifthenelse{\boolean{longversion}}{
\clearpage
\appendix
\section{Detailed Discussion of The Experiment}
\label{appx:detail_experiment}

We show some of the benchmarks from Termination Competition 2020 that are solved by \muval{} but not by \aprove{}, \ultimateautomizer{}, and \irankfinder{}.
We also discuss the strength of our method.

\paragraph{A benchmark containing a nonlinear operation.}
Fig.~\ref{fig:bench_nonlinear} shows one of the benchmarks that contains a nonlinear operation \texttt{y = y * y} but admits a linear ranking function.
Although handling nonlinear operation is difficult in general, \muval{} was able to verify termination.
The reason can be understood as follows.
(1) Our validator can find transition examples for this benchmark thanks to the recent development of SMT solvers.
(2) Our synthesizer can work as usual because it is example-based.

\paragraph{A benchmark that requires a long lasso-shaped non-terminating trace to prove non-termination.}
Fig.~\ref{fig:bench_long_lasso} is a benchmark that is non-terminating.
To prove non-termination of this benchmark, we need to find a long lasso-shaped trace that ends in the self loop when $\texttt{i} = \texttt{range} = 0$.
Finding such a long lasso is difficult in general, but \muval{} was able to find one in this benchmark: during CEGIS iterations, \muval{} found the self loop (i.e.\ explicit cycle) and then collected transition examples that were needed to prove reachability of cycles.

Our method can naturally collect transition examples ``backward'' from the self loop (an explicit cycle).
At the same time, our method collects transition examples ``forward'' from an initial state (in the benchmark of Fig.~\ref{fig:bench_long_lasso}, the pair of $\texttt{i} = 10$ and $\texttt{range} = 20$ is an initial state of the while loop).
When transition examples that are collected backward and forward form a trace to the self loop, our method notice that the benchmark is non-terminating.

\begin{figure}[tbp]
	\centering
	\begin{minipage}{0.45\textwidth}
		\begin{lstlisting}[language={C},basicstyle={\scriptsize\ttfamily},tabsize=2]
int main() {
	int x = ?;
	int y = 2;
	int res = 1;
	if (x < 0 || y < 1) { }
	else {
		while (x > y) {
			y = y*y;
			res = 2*res;
		}
	}
}
		\end{lstlisting}
		\subcaption{A benchmark containing a nonlinear operation.}
		\label{fig:bench_nonlinear}
	\end{minipage}
	\begin{minipage}{0.45\textwidth}
		\begin{lstlisting}[language={C},basicstyle={\scriptsize\ttfamily},tabsize=2]
int main() {
	int i = ?;
	int range = 20;
	while (0 <= i && i <= range) {
		if (!(0 == i && i == range)) {
			if (i == range) {
				i = 0;
				range = range-1;
			} else {
				i = i+1;
			}
		}
	}
}
		\end{lstlisting}
		\subcaption{A benchmark that requires a long lasso to prove non-termination}
		\label{fig:bench_long_lasso}
	\end{minipage}
	\caption{Some of the benchmarks}
\end{figure}

}{}

\end{document}